\documentclass[a4paper]{article}

\usepackage[abspath]{currfile}
\usepackage{graphicx}
\usepackage{amsmath}
\usepackage{mathtools}
\usepackage{amsopn}
\usepackage{amsfonts}
\usepackage{amssymb}
\usepackage{amsthm}
\usepackage{xspace}

\usepackage{subcaption}
\usepackage{makecell}

\usepackage{multirow}
\usepackage{booktabs}
\usepackage{ifthen}

\newcommand{\Params}{\mathcal{P}}

\newcommand{\FOMML}{\texttt{FOM-ML}\xspace}
\newcommand{\PODML}{\texttt{POD-ML}\xspace}
\newcommand{\PODDEIMML}{\texttt{POD-DEIM-ML}\xspace}

\newcommand{\NSRelErrState}{e_{\text{state}}^{\text{rel}}}
\newcommand{\NSRelErrDrag}{e_{\text{drag}}^{\text{rel}}}
\newcommand{\NSRelErrLift}{e_{\text{lift}}^{\text{rel}}}
\newcommand{\NSRelErrMaxState}{e_{\text{state}}^{\text{rel},\infty}}
\newcommand{\NSRelErrMaxDrag}{e_{\text{drag}}^{\text{rel},\infty}}
\newcommand{\NSRelErrMaxLift}{e_{\text{lift}}^{\text{rel},\infty}}

\usepackage{minted}

\newcommand{\Library}[1]{#1\xspace}

\newcommand{\pyMOR}{\Library{pyMOR}}
\newcommand{\NumPy}{\Library{NumPy}}
\newcommand{\SciPy}{\Library{SciPy}}
\newcommand{\FEniCS}{\Library{FEniCS}}
\newcommand{\FEniCSx}{\Library{FEniCSx}}
\newcommand{\ngsolve}{\Library{NGSolve}}
\newcommand{\dealii}{\Library{deal.II}}
\newcommand{\Python}{\Library{Python}}

\newcommand{\ScikitLearn}{\Library{scikit-learn}}
\newcommand{\PyTorch}{\Library{PyTorch}}

\newminted{python}{linenos,fontsize=\small}
\newmintinline[PY]{python}{}

\newcommand{\PyClass}[1]{\PY{#1}}
\newcommand{\PyClasses}[1]{\PY{#1}s}

\newcommand{\VectorArray}{\PyClass{VectorArray}\xspace}
\newcommand{\VectorArrays}{\PyClasses{VectorArray}\xspace}
\newcommand{\VectorSpace}{\PyClass{VectorSpace}\xspace}

\newcommand{\Operator}{\PyClass{Operator}\xspace}
\newcommand{\Operators}{\PyClasses{Operator}\xspace}
\newcommand{\Model}{\PyClass{Model}\xspace}
\newcommand{\Models}{\PyClasses{Model}\xspace}
\newcommand{\Reductor}{\PyClass{Reductor}\xspace}
\newcommand{\Reductors}{\PyClasses{Reductor}\xspace}

\newcommand{\C}{\mathbb{C}}
\newcommand{\R}{\mathbb{R}}

\newcommand{\deta}{\,\textnormal{d}\eta}
\newcommand{\ds}{\,\textnormal{d}s}
\DeclareMathOperator{\dif}{d\!}

\DeclareMathOperator{\colspan}{colspan}

\DeclareMathOperator*{\argmin}{arg\,min}
\DeclarePairedDelimiter{\Paren}{\lparen}{\rparen}
\DeclarePairedDelimiter{\Norm}{\lVert}{\rVert}
\DeclarePairedDelimiterX\set[1]\lbrace\rbrace{\def\given{\;\delimsize\vert\;}#1}
\DeclarePairedDelimiterXPP{\Diag}[1]{\operatorname{diag}}{\lparen}{\rparen}{}{#1}

\newcommand{\Laplace}{\mathcal{L}}
\newcommand{\Hinf}{\mathcal{H}_{\infty}}
\newcommand{\Htwo}{\mathcal{H}_{2}}
\newcommand{\Linf}{\mathcal{L}_{\infty}}
\newcommand{\Ltwo}{\mathcal{L}_{2}}

\newcommand{\T}{\top}

\usepackage{geometry}
\usepackage{orcidlink}

\usepackage{xspace}

\usepackage{minted}

\usepackage[capitalize]{cleveref}

\usepackage[title]{appendix}%
\usepackage{textcomp}%
\usepackage{manyfoot}%

\usepackage[backend=biber, style=numeric]{biblatex}
\theoremstyle{thmstyleone}%
\theoremstyle{thmstyletwo}%
\newtheorem{remark}{Remark}%

\theoremstyle{thmstylethree}%

\crefformat{equation}{(#2#1#3)}
\crefrangeformat{equation}{(#3#1#4) to~(#5#2#6)}
\crefmultiformat{equation}{(#2#1#3)}{ and~(#2#1#3)}{, (#2#1#3)}{ and~(#2#1#3)}

\title{Data-Driven Model Order Reduction with pyMOR}
\author{Hendrik Kleikamp\,\orcidlink{0000-0003-1264-5941}\,\thanks{IDea\_Lab, University of Graz, Leechgasse~34, 8010~Graz, Austria, \texttt{hendrik.kleikamp@uni-graz.at}.} \and
Petar Mlinarić\,\orcidlink{0000-0002-9437-7698}\,\thanks{University of Zagreb Faculty of Science, Department of Mathematics, Bijenička~cesta~30, 10000~Zagreb, Croatia, \texttt{petar.mlinaric@math.hr}.} \and
Stephan Rave\,\orcidlink{0000-0003-0439-7212}\,\thanks{Mathematics Münster, University of Münster, Einsteinstrasse~62, 48149~Münster, Germany, \texttt{stephan.rave@uni-muenster.de}.}\textsuperscript{\, ,\,}\thanks{Corresponding author, email: {\tt stephan.rave@uni-muenster.de}} \and
Felix Schindler\,\orcidlink{0000-0003-1582-7118}\,\thanks{Arup Deutschland GmbH, Speditionsstraße~9, 40221~Düsseldorf, Germany, \texttt{felix.schindler@arup.com}.}}

\date{July 29, 2026}

\begin{document}

\maketitle

\begin{abstract}
\pyMOR is a free and open-source software library of model order reduction algorithms for the
Python programming language.
Designed with classical model-based reduction methods for large-scale parametric partial differential
equation problems in mind, algorithms in \pyMOR are implemented in terms of operations on abstract
\VectorArray, \Operator and \Model interfaces, allowing for a seamless integration with external
solver codes implementing the full-order model.
For cases where a tight integration with the full-order model code is not feasible,
data-driven model order reduction algorithms, which only require simulation or measurement data of the
full-order model, are an attractive alternative.
In this work we discuss the data-driven methods that have been recently added to \pyMOR, show practical
examples of their application using \pyMOR and compare their performance with classical model-based methods.
We show that \pyMOR serves as a unified framework for combining model-based and data-driven methods,
enabling the construction of flexible and efficient hierarchical model reduction pipelines.
\end{abstract}


\maketitle

\section{Introduction}\label{sec:introduction}
In many areas of computational science and engineering, mathematical models based on partial differential equations~(PDEs) or control systems play a crucial role.
Typically, the solution of these equations requires high-dimensional discretizations or small time steps, leading to large computational costs.
These models often depend on certain physical and geometrical parameters or external controls.
In such cases, solving the parametric~PDE system for many different values of the parameters or simulating the control system for various different controls becomes computationally infeasible.
Model order reduction~(MOR) addresses this issue by replacing a given high-fidelity full-order model~(FOM) with a suitable surrogate reduced-order model (ROM).
We refer to~\cite{hesthaven2016certified,QuarteroniManzoniEtAl2016ReducedBasisMethods,haasdonk2017chapter} for introductions to parametric~MOR techniques and to~\cite{Ant05,AntoulasBeattieEtAl2020InterpolatoryMethodsModel,BennerSchildersEtAl2020ModelOrderReduction1} for methods addressing control systems.
\par
\pyMOR is a free and open source \Python library of MOR algorithms, that is designed for easily integrating MOR into existing computational science and engineering applications~\cite{ohlberger2014model,feinauer2019multibat}.
\pyMOR's algorithms are implemented in terms of operations on \VectorArray, \Operator and \Model objects\footnote{Monospace font indicates Python code samples and names from \pyMOR's API.}, which correspond to data structures and algorithms in external solver code implementing the FOM.
Compared to directly adding MOR code to the FOM solver, this approach allows implementing reusable MOR algorithms in an accessible scripting language, while relying on existing high-performance implementations for all high-dimensional FOM operations~\cite{milk2016pymor}.
Originally developed as a software for reduced basis (RB)~MOR of parametric~PDEs, a large variety of methods for control systems, such as balanced truncation (BT) or the iterative rational Krylov algorithm (IRKA) have been added to \pyMOR over the years~\cite{balicki2019systemtheoretic,mlinaric2021parametric}.
All these methods rely on projecting the dynamics of the FOM (expressed as \Operator equations defined by the full-order \Model) onto suitable low-dimensional subspaces represented by appropriate basis \VectorArrays.
\par
While classical FOM-based MOR has been successfully employed in many applications~\cite{BennerSchildersEtAl2020ModelOrderReduction}, it is intrusive in the sense that the MOR code needs to access internal data and algorithms (system matrices, nonlinear operators, domain-specific (non-)linear solvers) of the FOM solver.
In the case of \pyMOR, this usually amounts to exposing these objects via \Python bindings and writing small adapter classes to match \pyMOR's interfaces~\cite{milk2016pymor,TutorialBindingExternal}.
Although this approach is also beneficial for other applications and many open-source PDE solver libraries already include \Python bindings, there are still situations where an intrusive MOR approach is not viable.
In particular, this is often the case for commercial FOM solvers or situations where only measurement data of a physical system is available, but no full mathematical model of its dynamics.
\par
In recent years, data-driven methods gained significant attention in the MOR community.
These methods aim to construct ROM surrogates purely from FOM simulations or measurement data and without explicit knowledge of the underlying model.
A classical example for such a method in the context of parametric problems was proposed in~\cite{hesthaven2018nonintrusive}, where the authors suggest to build a RB using proper orthogonal decomposition~(POD)~\cite{Sirovich1987TurbulenceDynamicsCoherent} and afterwards learn the map from parameter to reduced coefficients by training a neural network.
On the other hand, methods such as the Loewner framework~\cite{MayA07}, the parametric adaptive Anderson-Antoulas~(p-AAA)~algorithm~\cite{rodriguez2023paaa} or the realization-independent iterative rational Krylov algorithm~(TF-IRKA)~\cite{beattie2012realization} are well-known examples of data-driven approaches to approximate transfer functions of (parametric) linear time-invariant dynamical systems.
Furthermore, system-identification methods such as the dynamic mode decomposition~(DMD)~\cite{schmid2010dynamic} or the eigensystem realization algorithm~(ERA)~\cite{juang1985eigensystem} constitute important approaches for constructing low-order models based purely on state-space or input-output data.
All these methods have been recently implemented in~\pyMOR.
\par
Over the years, various MOR software libraries have been developed.
We are aware of the following actively maintained open-source codes:
\Library{ModelOrderReduction.jl}~\cite{ModelOrderReductionJl},
\Library{ITHACA-FV}~\cite{2026ITHACAFVITHACAFV} and \Library{RBniCS/RBniCSx}~\cite{rozza2024real,2026RBniCSRBniCSx}
focus on model-based methods that rely on state-space approximations like POD or RB methods.
Each of these libraries is tightly coupled to a specific PDE solver ecosystem, however.
\Library{Pressio}~\cite{RizziBloniganEtAl2021PressioEnablingProjectionbased} offers abstractions similar to \pyMOR
for integrating large-scale PDE solvers, but it is also limited to state-space approximation methods.
Using \Library{C++} as implementation language, it is targeted at HPC environments,
with the downside that extending the library is challenging for MOR researchers lacking deep knowledge of \Library{C++}.
For control systems, \Library{MORLAB}~\cite{benner2023morlab} implements a large selection of
matrix-equation-based methods in \Library{MATLAB}.
Regarding data-driven MOR methods, we mention the \Python packages
\Library{EZyRB}~\cite{DemoTezzeleEtAl2018EZyRBEasyReduced},
\Library{PyDMD}~\cite{ichinaga2024pydmd},
\Library{PySPOD}~\cite{MengaldoMaulik2021PySPODPythonPackage},
\Library{PySINDy}~\cite{KaptanogluSilvaEtAl2022PySINDyComprehensivePython},
\Library{OpInf}~\cite{2026OperatorinferenceOpinf} and
\Library{pyNIROM}~\cite{dutta2021pynirom}, which each specialize in specific types of methods.
Finally we mention \Library{libROM}~\cite{librom}, which mainly implements basis generation and hyperreduction
algorithms for state-space approximation methods, but also implements the data-driven DMD algorithm.
Unlike \Library{Pressio} and \pyMOR, it does not provide any interfaces for integrating external solvers but relies
on user code to construct the ROM from the generated data.
\par
In summary, to the best of our knowledge, \pyMOR is the only software library that provides such an extensive selection
of model-based and data-driven MOR algorithms in a unified framework, addressing both control systems and parameterized PDE
problems in weak formulation.
\pyMOR's interface-based design allows applying these methods to both simple academic test cases and
real-world problems involving specialized high-dimensional discretizations or distributed solvers.
\par
In this work, we give an overview of the new data-driven MOR methods that are available in \pyMOR 2025.2, including short
mathematical descriptions and a discussion of their implementation in \pyMOR.
In multiple numerical experiments, we show how \pyMOR can be used as a unifying MOR framework for easily
exploring and comparing both data-driven and classical model-based MOR methods.
Further, we show how \pyMOR makes it possible to build hybrid hierarchical MOR pipelines that combine
model-based ROMs as FOM surrogates for generating training data for even more efficient data-driven
ROMs.
\par
The paper is organized as follows: In~\Cref{sec:general-structure}, we describe the core design of the software and discuss the most important building blocks of~\pyMOR.
In \cref{sec:models}, we introduce the classes of FOMs considered in this work and discuss their
realization as \pyMOR \Models.
Some of the model-based MOR methods implemented in \pyMOR are discussed in~\Cref{sec:model-based-methods}.
These traditional methods serve as a baseline for evaluating the data-driven approaches introduced in~\Cref{sec:data-driven-methods}.
In~\Cref{sec:experiments}, we present numerical results for the methods discussed in this paper.
We also provide excerpts of the \Python code used to conduct the numerical experiments.
We close with an outlook to future developments in~\pyMOR in~\Cref{sec:outlook}.

\section{General structure and main components of the software}\label{sec:general-structure}

\begin{figure}[htbp]
	\centering
	\resizebox{\linewidth}{!}{%
		\includegraphics{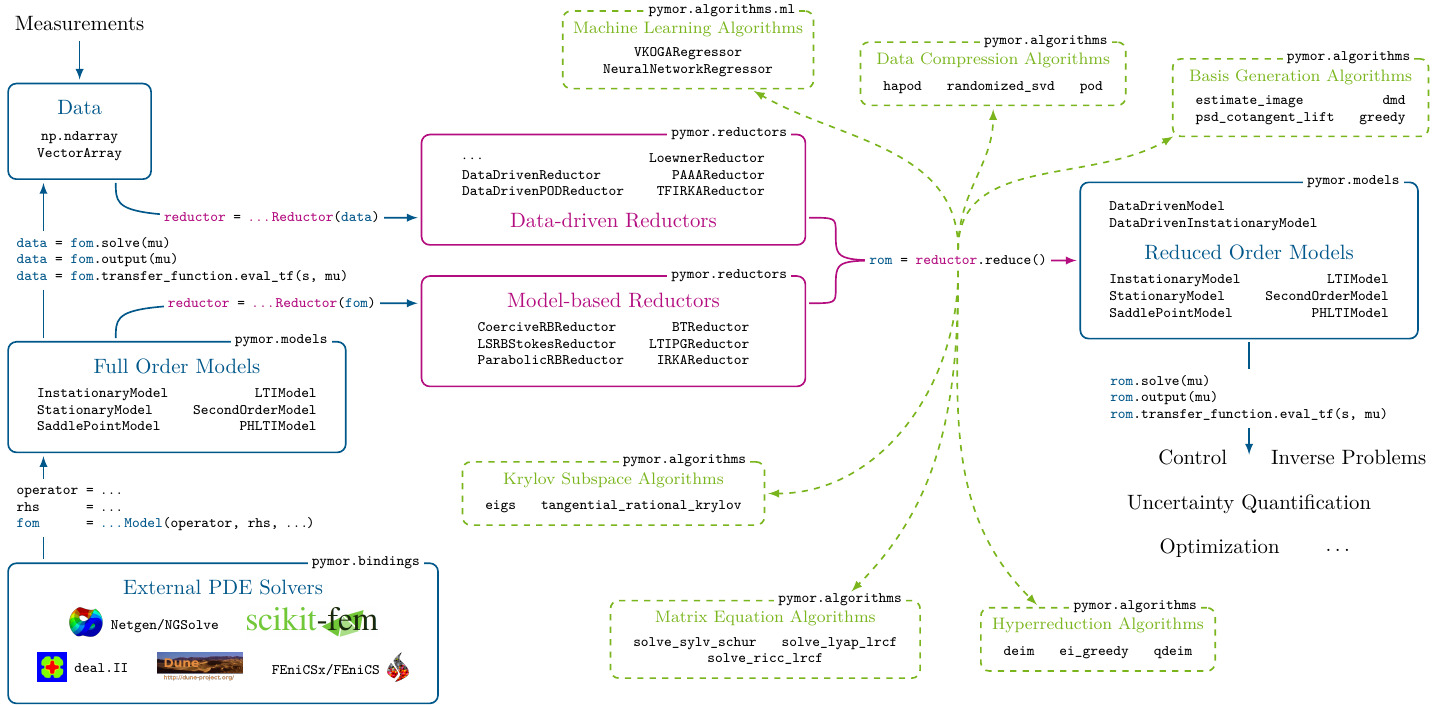}%
	}
	\caption{Interplay of selected~\pyMOR components (boxes), algorithms (dashed boxes) and user code.
		Arrows (with representative code snippets) indicate data flow, dashed arrows indicate ``makes use of''.
    }
	\label{fig:architecture-diagram-full}
\end{figure}

\pyMOR follows an object-oriented design, where the given FOM is represented by instances of \Model subclasses.
\Reductor objects operate on these FOM \Models to produce ROM \Models via the \PY{reduce} method (see also \cref{fig:architecture-diagram-full}).
\par
Each \Model class corresponds to a specific type of system of equations that constitutes the FOM or ROM.
For instance, \PY{StationaryModel} represents general (non-)linear systems~$A(x) = b$,
whereas \PY{LTIModel} represents linear, time-invariant control systems of the form~\cref{eq:lti_fom}.
The mathematical objects appearing in these equations are given by corresponding attributes of the \Model.
Both matrices and general nonlinear operators are represented by \Operator objects,
which can be applied to \VectorArrays, ordered list of vectors of same dimension.
In addition to encoding the mathematical structure of a model and holding the associated data, \Models
implement several methods to compute various quantities of interest, such as system response, state-space
trajectories or parameter sensitivities.
These methods are used both by \pyMOR itself in the reduction process, e.g., to generate snapshot data for
reduced basis construction, or by the user to integrate the generated ROM into downstream applications
like optimal control or uncertainty quantification.
\par
\pyMOR provides implementations of \Operators and \VectorArrays based on the \NumPy/\SciPy ecosystem,
which are also used in the ROMs generated by \pyMOR.
When the FOM is provided by an external solver, adapter classes need to be available, which match \pyMOR's
\Operator and \VectorArray interfaces to the corresponding structures in the solver.
Bindings for the open-source libraries \dealii, \FEniCS/\FEniCSx and \ngsolve are provided in the
\PY{pymor.bindings} package\footnote{In the case of \dealii, the \Library{pymor-dealii} package has to be
installed.}.
As \Models, either \pyMOR's built-in classes can be used, which use generic time steppers and the (non-) linear
\PyClasses{Solver} attached to the \Operators to simulate the model, or the user subclasses such a \Model
to directly call into the external solver's simulation routines.
\par
The \PY{pymor.algorithms} package contains various basic numerical algorithms that are required for the
MOR methods in \pyMOR.
For instance, the package includes algorithms for Gram-Schmidt orthonormalization,
(randomized) truncated SVD, Krylov subspace algorithms or low-rank matrix equation solvers.
By using \pyMOR's \Operator and \VectorArray abstractions, these algorithms can be used with any
supported linear algebra backend without requiring dedicated support for these operations in the external
solver.
\par
In the case of data-driven MOR, \Reductors no longer operate on input FOMs.
Instead, they require simulation or measurement data associated with the FOM.
This data is provided to the \Reductor either as a \VectorArray or as a \NumPy array, depending on
whether the data is related to the state or output of the FOM.
When a FOM is available, the FOM's interface methods can be used to generate the needed input data.
This makes it possible to easily compare data-driven to classical model-based MOR approaches.
\par
In~\cref{fig:architecture-diagram-full}, we visualize a typical MOR workflow using \pyMOR.
For a more detailed discussion of \pyMOR's architecture, including further design aspects such as
immutability of \Operators and \Models, copy-on-write semantics for \VectorArrays, multiple dispatch
via \PyClasses{Ruletable} or parallelization, we refer to~\cite{milk2016pymor} and to
\pyMOR's extensive online documentation\footnote{\url{https://docs.pymor.org}}.

\section{Full-order models}\label{sec:models}

In this section, we introduce the different classes of FOMs for which we discuss MOR in this paper.
We begin by introducing general nonlinear control systems in \cref{sec:nonlinear_control_systems}.
An important special case are linear time-invariant systems (\cref{sec:lti_systems}),
for which a plethora of specialized methods have been developed over the years.
In the numerical experiments, we will discuss a Navier-Stokes model that uses an IMEX method
for time discretization.
In order to be able to exploit this additional structure for MOR, we introduce a specialized
IMEX model in \cref{sec:imex_models}.

\subsection{Nonlinear control systems}\label{sec:nonlinear_control_systems}
All models that we consider in this paper can be regarded as instances of
parametric control systems of the form
\begin{equation}
    \label{eq:general_fom}
    \begin{aligned}
        E \dot x(t) &= f(t, x(t), u(t), \mu), & t \in [0,T], \\
        x(0) &= 0, \\
        y(t) &= g(t, x(t), u(t), \mu), & t \in [0,T].
    \end{aligned}
\end{equation}
Here, $\mu \in \Params$ is a given parameter from an arbitrary set of parameters~$\Params \subset \R^P$,
$u\colon [0,T] \to \R^m$ is a given input function, $x\colon [0, T] \to \R^n$ is the corresponding
state-space solution determined by~$f\colon \R \times \R^n \times \R^m \times \Params \to \R^n$, $E \in \R^{n\times n}$, and~$y\colon [0, T] \to \R^p$ is the corresponding output determined by~$g\colon \R \times \R^n \times \R^m \times \Params \to \R^p$.
In our notation we neglect the dependence of~$x$ and~$y$ on~$\mu$ and~$u$.
In \pyMOR, the system \cref{eq:general_fom} is represented by the \PyClass{InstationaryModel} class,
where $E$, $-f$ and $g$ are given by the \PY{mass}, \PY{operator} and \PY{output_functional} attributes.

Not every system~\Cref{eq:general_fom} must have a control input~$u(t)$ and a parameterization~$\mu$.
While MOR methods targeted at control systems are generally quite different to methods
for parameterized problems, we note that one could interpret~$u(t)$ as a time-dependent parameter or~$\mu$ as
the coefficients of an input-function w.r.t.\ a finite-dimensional basis of some function space.
Indeed, \pyMOR internally represents both the components of~$\mu$ and~$u(t)$ as items of the same
\PY{Mu} object of time-dependent parameter values, which are passed to~$f$ whenever it is applied.

Often the system~\Cref{eq:general_fom} has additional structure which we might exploit for MOR.
For instance, $f$ might be linear, or the system might have a block structure arising
from a second-order differential equation.
A particularly important case for parametric MOR is a parameter-separable~$f$ of the form
\begin{equation}
    \label{eq:parameter_separable_general}
    f(t, x, u, \mu) = \sum_{q=1}^Q\theta_q(t, u, \mu) f_q(x),
\end{equation}
where~$f$ is a parametric linear combination of non-parametric operators~$f_q\colon\R^n \to \R^n$.
This structure is encoded in \pyMOR by defining~$f$ as an instance of \PyClass{LincombOperator}, which holds
the\\\PyClasses{ParameterFunctional} $\theta_q$ and the \Operators~$f_q$ as attributes.
\par
While we consider only ODE systems here, note that~\Cref{eq:general_fom} might arise from the discretization of a
PDE. In this case it might be relevant to consider appropriate norms and inner products on~$\R^n$
and to distinguish Hilbert spaces from their dual spaces (in case~$f$ comes from a weak formulation).
\pyMOR has extensive support for this case, for instance by providing inner~\PY{product} arguments for
many algorithms or by supporting the computation of Riesz representatives.

\subsection{Linear time-invariant systems}\label{sec:lti_systems}
In the following section we introduce the special case of non-parametric linear
time-invariant (LTI) systems~\cite{Ant05, BenOCetal17, BreitenStykel2021BT, BennerFeng2021MM},
where~$f$ and~$g$ in \cref{eq:general_fom} are given by $f(t, x, u, \mu) = Ax + Bu$ and $g(t, x, u, \mu) =
Cx + Du$.
In other words, we consider systems of the form:
\begin{equation}
    \label{eq:lti_fom}
    \begin{aligned}
        E\dot{x}(t) &= A x(t) + B u(t), & t \in [0,T],\\
        x(0) &= 0, \\
        y(t) &= C x(t) + D u(t), & t \in [0,T].
    \end{aligned}
\end{equation}
In \pyMOR, such systems are represented by the \PyClass{LTIModel} class with attributes \PY{A}, \PY{B},
\PY{C}, \PY{D} and \PY{E}.
\par
We assume that~$E$ is invertible and~$E^{-1} A$ is Hurwitz, i.e.,
all of its eigenvalues are in the open left half-plane.
Therefore, the system~\eqref{eq:lti_fom} is asymptotically stable.
\par
System-theoretic methods have the goal of approximating the input-to-output
mapping~$u \mapsto y$ (an infinite-dimensional linear operator),
which is given explicitly as
\begin{equation*}
    y(t)
    =
    \int_0^t C e^{\tau E^{-1} A} E^{-1} B u(t - \tau) \dif{\tau}
    + D u(t),
\end{equation*}
i.e., the output is given by a convolution
\begin{equation*}
    y = h * u, \quad
    h(t) =
    \begin{cases}
        C e^{t E^{-1} A} E^{-1} B + \delta(t) D, & t \ge 0, \\
        0, & t < 0,
    \end{cases}
\end{equation*}
with~$h$ the impulse response and~$\delta$ the Dirac delta distribution.
Considering the frequency domain simplifies the convolution to the standard
multiplication, specifically,
if~$\Laplace$ is the Laplace transform and~$U = \Laplace(u)$, $Y = \Laplace(y)$, and~$H = \Laplace(h)$ exist,
then
\begin{equation*}
    Y(s) = H(s) U(s), \quad
    H(s) = C (s E - A)^{-1} B + D,
\end{equation*}
with~$H$ the transfer function.
Note that~$H$ is a matrix-valued rational function
that is strictly proper if~$D = 0$ and proper otherwise.
\par
Many system-theoretic methods then focus on finding a~ROM with a transfer
function~$\hat{H}$ such that~$H - \hat{H}$ is small in some norm.
Two popular norms are the~$\Hinf$ and~$\Htwo$ norms,
\begin{equation*}
    \Norm{H - \hat{H}}_{\Hinf}
    =
    \sup_{\omega \in \R} \,
    \Norm{H(i \omega) - \hat{H}(i \omega)}_2,
    \quad
    \Norm{H - \hat{H}}_{\Htwo}
    =
    \Paren*{
        \frac{1}{2 \pi}
        \int_{-\infty}^{\infty}
        \Norm{H(i \omega) - \hat{H}(i \omega)}_F^2
        \dif{\omega}
    }^{1/2}
\end{equation*}
which satisfy output error bounds
\begin{equation*}
    \Norm{y - \hat{y}}_{\Ltwo}
    \le
    \Norm{H - \hat{H}}_{\Hinf}
    \Norm{u}_{\Ltwo},
    \quad
    \Norm{y - \hat{y}}_{\Linf}
    \le
    \Norm{H - \hat{H}}_{\Htwo}
    \Norm{u}_{\Ltwo}.
\end{equation*}
Therefore, for a fixed input of bounded~$\Ltwo$ norm,
making either~$\Norm{H - \hat{H}}_{\Hinf}$ or~$\Norm{H - \hat{H}}_{\Htwo}$ small
also makes the output error~$y - \hat{y}$ small (in a corresponding norm).
Note that the $\Htwo$ norm is bounded only for rational functions that are
strictly proper,
while the $\Hinf$ norm is bounded for all proper rational functions.
\par
\pyMOR allows to compute $y(t)$ for arbitrary input functions $u(t)$ via the \Model's \PY{output}
method.
The transfer function of an \PyClass{LTIModel} is available as the \PY{transfer_function} attribute.
The~$\Hinf$ and~$\Htwo$ norms of the model can be determined using the \PY{hinf_norm} and \PY{h2_norm} methods.
\par
It is worth noting that the transfer function~$H(s) = C (s E - A)^{-1} B$ can be
viewed as an output of a parametric stationary problem
\begin{equation}\label{eq:tf_fom}
    \begin{aligned}
        (s E - A) X(s) &= B, \\
        H(s) &= C X(s),
    \end{aligned}
\end{equation}
where~$s$ is considered as a parameter.
Therefore, parametric MOR methods for stationary problems, such as RB methods,
can be applied in the setting of~LTI systems (see, e.g.,~\cite{HesB13,MliG23}).
The system~\cref{eq:tf_fom} can be represented in \pyMOR as a \PyClass{StationaryModel}
with $sE - A$ as \PY{operator}, $B$ as \PY{rhs} and $C$ as \PY{output_functional} attributes.
\par
We will also consider linear systems with additional structure.
Second-order systems of the form
\begin{equation}
    \label{eq:so_fom}
    \begin{aligned}
        M \ddot{x}(t) + E \dot{x}(t) + K x(t) &= B u(t), & \quad
        x(0) = 0,\ \dot{x}(0) = 0, \\
        y(t) &= C_p x(t) + C_v \dot{x}(t), &
    \end{aligned}
\end{equation}
are represented in \pyMOR by the \PyClass{SecondOrderModel} class.
Viewing $\dot{x}$ as an additional variable, each such system can be transformed into a first-order LTI
system \cref{eq:lti_fom} using the \PY{to_lti} method.
Port-Hamiltonian systems are passive LTI system with a state-space realization of the form
\begin{equation}
    \label{eq:ph_fom}
    \begin{aligned}
        E \dot{x}(t) &= (J - R) Q x(t) + G u(t), \quad
        x(0) = 0, \\
        y(t) &= G^\T Q x(t),
    \end{aligned}
\end{equation}
where~$Q^\T E$ is symmetric positive definite, $J$ is skew-symmetric, and~$R$
is symmetric positive semidefinite
(in the strictly proper case).
In \pyMOR, such systems are represented by the \PyClass{PHLTIModel} subclass of \PyClass{LTIModel}.

\subsection{IMEX models}\label{sec:imex_models}
In view of the Navier-Stokes example discussed in ~\Cref{sec:navier_stokes}, we also consider
discrete-time models using IMEX time stepping of the form
\begin{equation}
    \label{eq:imex}
    \begin{aligned}
        (E - \Delta t\cdot F_i(\mu)) x^{(k+1)} &= E x^{(k)} + \Delta t\cdot f_e(t_k, x^{(k)}, u^{(k)}, \mu), \\
        x^{(0)} &= 0.
    \end{aligned}
\end{equation}
Here, $x^{(k)}, u^{(k)}$ denote the approximate solution and the input at time~$t_k \coloneqq  k \cdot
\Delta t$, and we assume a splitting of~$f$ in \cref{eq:general_fom} into a linear, time- and input-invariant implicit part~$F_i$ and
an explicit part~$f_e$:
\begin{align*}
    f(t, x, u, \mu) = F_i(\mu) x + f_e(t, x, u, \mu).
\end{align*}

In order to make use of this structure for MOR, we need to expose it to~\pyMOR.
As~\pyMOR currently does not include a \Model describing IMEX discrizations of this form,
we define a custom \Model, which has~$E$, $F_i$, $f_e$ as \Operator attributes:

\begin{pythoncode}
class IMEXModel(Model):

    def __init__(self, E, F_i, f_e, g, T, dt, solver=None, visualizer=None):
        super().__init__(visualizer=visualizer)
        self.__auto_init(locals())
        self.solution_space = E.source
        self.parameters_internal = {'t': 1}
        self.dim_output = g.range.dim
\end{pythoncode}

In line~5, we save all~\PY{__init__} arguments as instance attributes with the same name using~\pyMOR's
\PY{__auto_init} idiom.
Next we infer the \VectorSpace of the state-space solutions of the \Model from~\PY{E}.
Specifying time as an internal parameter in line~7 informs~\pyMOR that the time parameter will be set
internally by the \Model, so its interface methods (\PY{solve}, \PY{output}, etc.) will not require time
as an input parameter, even though some of the passed \Operators might be time-dependent.
\par
Since we want to use~\PyClass{IMEXModel} also as a~ROM, we need to implement the IMEX time stepping in
\PY{_compute} (which is invoked by the user facing~\PY{compute}, \PY{solve} or \PY{output} methods):
\begin{pythoncode}
    def _compute(self, quantities, data, mu=None):  # slightly simplified
        dt = self.dt
        num_steps = int(self.T / dt)
        X = self.solution_space.empty(reserve=num_steps)
        Y = np.zeros((self.dim_output, num_steps))

        g = self.g.assemble(mu)
        LHS = (self.E - dt * self.F_i).with_(solver=self.solver).assemble(mu)
        RHS = self.E + dt * self.f_e

        x = self.solution_space.zeros()
        for i, t in enumerate(np.arange(num_steps) * dt):
            r = RHS.apply(x, mu=mu.at_time(t))
            x = LHS.apply_inverse(r)
            X.append(x)
            Y[:, i] = g.apply(x).to_numpy().ravel()

        data['solution'] = X
        data['output'] = Y
\end{pythoncode}
Our implementation of~\PY{_compute} makes the assumption that~$E$, $F_i$ and~$g$ do not depend on time,
allowing us to pre-assemble the linear left-hand side of~\Cref{eq:imex} and the output functional once
before time stepping (lines 7 and 8).
We also specify that we want to use the~\PyClass{Solver} stored by the model to solve the update
equations of~\Cref{eq:imex}.

\section{Model-based MOR methods}\label{sec:model-based-methods}

We briefly discuss some of the model-based methods that are implemented in~\pyMOR.
By ``model-based'' we mean that these methods require, at least to some extent, access to a mathematical model for
the full-order dynamics that shall be reduced.

\subsection{Petrov-Galerkin projection}\label{sec:petrov-galerkin-projection}
All model-based MOR methods that we discuss are based on a Petrov-Galerkin projection of the~FOM.
This means that we are given projection matrices
\begin{align*}
    V, W \in \R^{n \times N},
\end{align*}
with~$N \ll n$ such that~$\colspan{V} \coloneqq  V_N \subset \R^n$ is an approximation space for~$x(t)$. Thus,
\begin{align*}
    x(t) \approx V \hat{x}(t)
\end{align*}
for some~$\hat{x}(t) \in \R^N$ (where $\hat{\cdot}$ indicates reduced quantities).
However, substituting this approximation into the nonlinear control system~\Cref{eq:general_fom} would yield an over-determined system,
so we multiply from the left with~$W^\T$ to obtain
\begin{equation}
    \label{eq:general_rom}
    \begin{aligned}
        \hat E \dot {\hat x}(t) &= \hat f(t, \hat{x}(t), u(t), \mu), & t \in [0,T], \\
        \hat{x}(0) &= 0, \\
        \hat{y}(t) &= \hat{g}(t, \hat{x}(t), u(t), \mu), & t \in [0,T],
    \end{aligned}
\end{equation}
where
\begin{equation}
    \label{eq:general_rom_quantities}
    \begin{aligned}
        \hat{E} &\coloneqq  W^\T  E  V, \\
        \hat{f}(t, \hat{x}, u, \mu) &\coloneqq  W^\T  f(t, V \hat{x}, u, \mu), \\
        \hat{g}(t, \hat{x}, u, \mu) &\coloneqq  \hphantom{{}W^\T  {}}g(t, V \hat{x}, u, \mu),
    \end{aligned}
\end{equation}
with~$\hat{E} \in \R^{N \times N}$, $\hat{f}\colon \R \times \R^N \times \R^m \times \Params \to \R^N$ and~$\hat{g}\colon \R \times \R^N \times \R^m \times \Params \to \R^p$ only acting on the reduced coordinates.
In~\pyMOR, we represent the tall and skinny matrices~$V, W$ as \VectorArrays, whereas~$f,g,E$ as
\Operators.
Petrov-Galerkin projection is performed using the~\PY{project} method:
\begin{pythoncode}
    E_hat = project(E, W, V)
    f_hat = project(f, W, V)
    g_hat = project(g, None, V)
\end{pythoncode}
Here, \PY{None} indicates that the identity is to be used as left-projection matrix.
\PY{project} automatically takes the structure of the given \Operators~\PY{E}, \PY{f}, \PY{g} into account.
For instance, assume that~$f$ is parameter-separable \cref{eq:parameter_separable_general} of the form
\begin{equation}
    \label{eq:parameter_separable}
    f(t, x, u, \mu) = \theta_0(t, u, \mu) F_0 x  + \theta_1(t, u, \mu) F_1 x,
\end{equation}
with matrices~$F_0, F_1 \in \R^{n \times n}$ and coefficient functions~$\theta_0, \theta_1\colon \R \times
\R^m \times \mathcal{P} \to \R$, we can write~$\hat f$ as
\begin{equation}
    \label{eq:projected_affine_decomposition}
    \hat f(t, \hat{x}, u, \mu) = \theta_0(t, u, \mu) \underbrace{(W^\T F_0 V)}_{\hat{F}_0} \hat{x}
        + \theta_1(t, u, \mu) \underbrace{(W^\T F_1 V)}_{\hat{F}_1} \hat{x}.
\end{equation}
This allows us to compute the matrix of~$\hat f(t, \cdot, u, \mu)$ by a linear combination of the
pre-computed~$N\times N$ matrices~$\hat{F}_0$ and~$\hat{F}_1$.
We can make~\pyMOR aware of this structure by writing
\begin{pythoncode}
    F = theta_0 * F_0 + theta_1 * F_1
\end{pythoncode}
where~\PY{F_0, F_1} are~\pyMOR \Operators and~\PY{theta_0, theta_1} are
\PyClasses{ParameterFunctional}.
The result is a~\PyClass{LincombOperator} encoding this decomposition.
\PY{project} will detect that~\PY{E} is a~\PyClass{LincombOperator} of non-parametric linear operators
and return a new~\PyClass{LincombOperator} according to~\Cref{eq:projected_affine_decomposition}.
\par
There are many different approaches to determine the projection matrices $V$ and $W$.
In snapshot-based MOR methods like RB methods, $V$ is computed from snapshots vectors~$x_1, \ldots, x_{n_\textnormal{train}}\in\R^n$ obtained from observations or~FOM simulations.
For parameterized problems, greedy algorithms, which iteratively add worst-approximated snapshot vectors
to the reduced basis, generate~$V$ with quasi-optimal approximation properties in the
sense of Kolmogorov~\cite{DeVorePetrovaEtAl2013GreedyAlgorithmsReduced}.
Alternatively, POD~(\cref{sec:pod}) can be used to determine $V$, as we will do in \cref{sec:navier_stokes}.
For elliptic or parabolic problems, it is often sufficient to let $W\coloneqq V$ (Galerkin projection).
For inf-sup stable problems, $W$ can be constructed to optimally stabilize the ROM for given $V$ (see, e.g.,
\cite{DahmenPleskenEtAl2014DoubleGreedyAlgorithms}).

\subsection{Least-squares residual minimization}
\label{sec:imex_lspg}
As in the continuous-time case, we can apply Petrov-Galerkin projection to the discrete-time IMEX
model \cref{eq:imex} to obtain the system
\begin{equation}
    \label{eq:pg_imex_rom}
    (\hat{E} - \Delta t \cdot \hat{F}_i(\mu)) \hat{x}^{(k+1)}
        = \hat{E} \hat{x}^{(k)} + \Delta t \cdot \hat{f}_e(t_k, \hat{x}^{(k)}, u^{(k)}, \mu).
\end{equation}
As an alternative to finding an appropriate left-projection matrix $W$, we consider the least-squares minimization problem
\begin{equation}
    \label{eq:least_squares_rom}
    \hat{x}^{(k+1)} \coloneqq  \argmin_{\hat{x} \in \R^N}
        \Norm*{(E - \Delta t\cdot F_i(\mu)) V \hat{x}
                - E V\hat{x}^{(k)} - \Delta t \cdot f_e(t_k, V\hat{x}^{(k)}, u^{(k)}, \mu)}^2.
\end{equation}
Using \pyMOR, we can solve \cref{eq:least_squares_rom} as follows:
\begin{pythoncode}
    LHS = project(E - dt * F_i, None, V).with_(solver=QRLeastSquaresSolver())
    RHS = project(E + dt * f_e, None, V)
    ...
    RHS_kp1 = RHS.apply(x_k, mu=mu.at_time(t_k))
    x_kp1   = LHS.apply_inverse(RHS_kp1, mu=mu)
\end{pythoncode}
Here, we multiply~\PY{E} and~\PY{F_i} only from the right, which leads to a tall and skinny system.
The resulting least-squares problem is solved using~\PyClass{QRLeastSquaresSolver}, which performs a QR
decomposition of the \VectorArray~\PY{RHS_kp1} using~\pyMOR's~\PY{gram_schmidt} algorithm.
As discussed, the input~$u$ is considered as a time dependent component of~\PY{mu}, which is
evaluated at~$t_k$ using the~\PY{at_time} method.
\par
While the resulting least-squares system is of low-rank, we still need to perform high-dimensional
computations in~$\R^n$ to solve it.
We can avoid this by finding a tall and skinny projection matrix~$W \in \R^{n \times M}$, $M \ll n$, such that~$\colspan E \subseteq \colspan W$.
If, in addition, the columns of~$W$ are orthonormal, then~\eqref{eq:least_squares_rom} is equivalent to
solving
\begin{align}
    \label{eq:projected_least_squares_rom}
    \hat{x}^{(k+1)} \coloneqq  \argmin_{\hat{x} \in \R^N}
        \Norm*{W^\T (E - \Delta t\cdot F_i(\mu)) V \hat{x} - W^\T E V\hat{x}^{(k)} - \Delta t \cdot W^\T f_e(t_k, V\hat{x}^{(k)}, u^{(k)}, \mu)}^2.
\end{align}
\pyMOR's~\PY{estimate_image} method can compute an appropriate~$W$ for several classes of \Operators,
including parameter-separable operators as in~\Cref{eq:parameter_separable}.
For solving~\Cref{eq:projected_least_squares_rom}, we would use the following code:
\begin{pythoncode}
    W = estimate_image([E, F_i], [], V)
    LHS = project(E - dt * F_i, W, V).with_(solver=ScipyQRLeastSquaresSolver())
    RHS = project(E + dt * f_e, W, V)
    ...
    RHS_kp1 = RHS.apply(x_k, mu=mu.at_time(t_k))
    x_kp1   = LHS.apply_inverse(RHS_kp1, mu=mu)
\end{pythoncode}
By projecting with~\PY{W}, we know that~\PY{LHS} will assemble to a low-dimensional
\PyClass{NumpyMatrixOperator}.
Thus, we can use a more efficient \SciPy-based solver instead of the generic
\PyClass{QRLeastSquaresSolver} which works with arbitrary \VectorArrays.

\begin{remark}
    \label{remark:petrov_galerkin_least_squares_equivalance}
    The least-squares problem~\Cref{eq:least_squares_rom} turns out to be equivalent to
    the Petrov-Galerkin projection~\Cref{eq:pg_imex_rom}, when we choose~$W$ to span the parameter-dependent test space
    \begin{align*}
        \colspan W = \colspan (E - \Delta t\cdot F_i(\mu))\cdot V.
    \end{align*}
    See~\cite{CarlbergBaroneEtAl2017GalerkinLeastsquaresPetrov} for a detailed discussion.
\end{remark}

\subsection{Empirical interpolation}\label{sec:deim}

Petrov-Galerkin projection or least-squares residual minimization both lead to a reduced-order
formulation of the model's dynamics in the reduced coordinates~$\hat{x}$.
However, when~$f$ is nonlinear, the evaluation of
\begin{align*}
    \hat{f}(t, \hat{x}, u, \mu) = W^\T f(t, V \hat{x}, u, \mu)
\end{align*}
in general still requires computing the high-dimensional vector~$V \hat{x} \in \R^n$, applying~$f$
to it and projecting back to~$\R^N$ by multiplying with~$W^\T$.
In most cases, the costs of evaluating~$\hat{f}$ will significantly limit the computational efficiency of
the~ROM.
\par
To overcome this issue, so-called hyperreduction techniques are applied.
We discuss here the discrete empirical interpolation method (DEIM)~\cite{BarraultMadayEtAl2004EmpiricalInterpolationMethod, HaasdonkOhlbergerEtAl2008ReducedBasisMethod, ChaturantabutSorensen2010NonlinearModelReduction},
where~$f$ is replaced by an
approximation~$\tilde{f}$ solving an interpolation problem of the form
\begin{align*}
    \tilde{f}(t, x, u, \mu) \in \colspan Z \quad\text{s.t.}\quad
    \tilde{f}(t, x, u, \mu)_i = f(t, x, u, \mu)_i \quad \text{for }i \in \mathcal{I}.
\end{align*}
The index set of interpolation points~$\mathcal{I}$ and the interpolation basis~$Z \in \R^{n\times M}$,
$M\coloneqq \lvert\mathcal{I}\rvert$ are computed (empirically) from given simulation data.
For the interpolation points, the EI-Greedy~\cite{BarraultMadayEtAl2004EmpiricalInterpolationMethod}
or more robust Q-DEIM~\cite{DrmacGugercin2016NewSelectionOperator} are used.
The interpolation basis can either be an existing projection basis, or be computed from evaluations of~$f$
using POD~\cite{ChaturantabutSorensen2010NonlinearModelReduction} or a
greedy search~\cite{HaasdonkOhlbergerEtAl2008ReducedBasisMethod}.
\par
Let~$i_1, \ldots, i_M$ be an arbitrary enumeration of~$\mathcal{I}$, and let~$S \in \R^{n\times M}$ be the matrix with entries~$S_{i,j} \coloneqq  \delta_{i,i_j}$.
Left-multiplication with~$S^\T$ selects the rows of~$f(t, x, u, \mu)$ with indices~$i_1, \ldots,
i_M$, so we can write~$\tilde{f}$ as
\begin{equation}
    \label{eq:deim_matrix_repr}
    \tilde{f}(t, x, u, \mu) = Z (S^\T Z)^{-1} S^\T f(t, x, u, \mu).
\end{equation}
Note that to evaluate~$\tilde{f}$, we only need to evaluate~$f$ at the~$M \ll n$ degrees of freedom
(DoFs)~$i_j$.
When~$f$ comes from a~PDE discretization with local stencil, this means that we only need to know~$x$
at some DoFs~$\mathcal{J}$, where~$M' = \lvert\mathcal{J}\rvert = \mathcal{O}(M)$.
Let~$l_1, \ldots, l_{M'}$ be an enumeration of~$\mathcal{J}$, and denote by~$S' \in R^{n \times M'}$
the matrix with entries~$S'_{i,j} \coloneqq  \delta_{i, l_j}$.
Then we can write
\begin{equation}
    \label{eq:deim_local_eval}
    S^\T  f(t, x, u, \mu) = f_r(t, S'^\T  x, u, \mu),
\end{equation}
where~$f_r(t, \cdot, u, \mu)\colon R^{M'} \to R^{M}$ is~$f$ restricted to~$i_1, \ldots, i_M$.
Applying Petrov-Galerkin projection to~$\tilde{f}$ given by~\Cref{eq:deim_matrix_repr} and substituting
\Cref{eq:deim_local_eval}, we obtain the hyper-reduced operator
\begin{equation}
    \label{eq:deim_projected}
    \hat{\tilde{f}}(t, \hat{x}, u, \mu) =
    (W^\T  Z)  (S^\T Z)^{-1} \cdot f_r(t, S'^\T V \hat{v}, u, \mu).
\end{equation}
After pre-computing the reduced-order matrices~$W^\T Z$, $S^\T Z$ and~$S'^\T V$,
we can evaluate~$\hat{\tilde{f}}$ with an effort of~$\mathcal{O}(NM + M^3 + M' + M'N)$,
independently of the~FOM order~$n$.
\par
In~\pyMOR, we can obtain~$\hat{\tilde{f}}$ as follows:
\begin{pythoncode}
    U = ... # compute f evaluations on training trajectories
    I, Z = qdeim(U, modes=M, pod=True)
    f_tilde = EmpiricalInterpolatedOperator(f, I, Z, triangular=False)
    f_tilde_hat = project(f_tilde, W, V)
\end{pythoncode}
For an efficient computation of~\PY{f_tilde}, \PyClass{EmpiricalInterpolatedOperator} needs to evaluate the
restricted operator~$f_r$.
To obtain this operator, \PyClass{EmpiricalInterpolatedOperator} internally calls
\begin{pythoncode}
    f_r, J = f.restricted(I)
\end{pythoncode}
which returns~$f_r$ together with the source DoFs~$\mathcal{J}$ needed for the evaluation of~$f_r$.
\par
As in the parameter-separable case, \PY{project} detects the structure of~\PY{f_tilde} and ensures that
the projected~\PyClass{Operator} can be efficiently evaluated according to~\Cref{eq:deim_projected}.
In this code snippet we prescribe the dimension~$M$ of the POD interpolation basis.
Alternatively we could have prescribed POD truncation error tolerances based on the singular value
decay of the snapshot matrix~\PY{U}.
Setting~\PY{pod=False} would have used~\PY{Z = U} as interpolation basis.
Using the~\PY{ei_greedy} algorithm instead of~\PY{qdeim} would have yielded a lower-triangular
interpolation matrix~$S^\T Z$.
In that case, passing~\PY{triangular=True} would allow~\PyClass{EmpiricalInterpolatedOperator} to choose
a faster solver for triangular matrices in its~\PY{apply} method.
\par
In \cref{sec:navier_stokes}, we will combine empirical interpolation with least-squares residual
minimization for the reduction of a Navier-Stokes FOM.

\subsection{Balanced truncation}

Balanced truncation (BT)~\cite{Moo81} is a MOR method for LTI systems \cref{eq:lti_fom}.
It consists of two steps:
\begin{enumerate}
\item balancing:
    finding a state-space transformation~$\tilde{x}(t) = T x(t)$,
    where~$T \in \R^{n \times n}$ is an invertible matrix,
    such that the new states~$\tilde{x}_1, \tilde{x}_2, \ldots, \tilde{x}_n$ are
    sorted from most to least ``important'',
\item truncation:
    removing the less important states and keeping only the first~$N$.
\end{enumerate}
The ``importance'' is measured in two parts:
input-to-state and state-to-output mapping.
The input-to-state importance is measured using the controllability Gramian
\begin{equation*}
    P
    =
    \int_0^\infty
    e^{t E^{-1} A} E^{-1} B
    B^\T E^{-\T} e^{t A^\T E^{-\T}}
    \dif{t},
\end{equation*}
while the state-to-output importance using the observability Gramian
\begin{equation*}
    E^\T Q E
    =
    \int_0^\infty
    e^{t A^\T E^{-\T}} C^\T
    C e^{t E^{-1} A}
    \dif{t}.
\end{equation*}
From~$P$ and~$E^\T Q E$, one can compute a balancing transformation~$T$,
or directly construct reduced basis matrices~$V$ and~$W$ to perform
projection-based reduction as described in~\Cref{sec:petrov-galerkin-projection}.
The~ROM obtained in this way satisfies an error bound based on
the eigenvalues~$\sigma_1, \sigma_2, \ldots, \sigma_n$ of the product~$E^\T Q E P$,
called the Hankel singular values.
If~$\sigma_N > \sigma_{N + 1}$,
then truncating to the first~$N$ states produces an asymptotically stable~ROM
satisfying the~$\Hinf$ error bound
\begin{eqnarray*}
    \Norm{H - \hat{H}}_{\Hinf}
    \le
    2
    \sum_{i = N + 1}^n
    \sigma_i.
\end{eqnarray*}
Computationally, one uses that~$P$ and~$Q$ satisfy Lyapunov equations
\begin{align*}
    A P E^\T
    + E P A^\T
    + B B^\T
    &= 0, \\
    A^\T Q E
    + E^\T Q A
    + C^\T C
    &= 0,
\end{align*}
for which there exist efficient methods to find low-rank approximations
$P \approx Z_P Z_P^\T$ and~$Q \approx Z_Q Z_Q^\T$,
several of which are available as solvers in ~\PY{pymor.algorithms.lyapunov}.
Also, instances of \PY{LTIModel} have a \PY{gramian} method that can compute
$P$ or $Q$ (although $Q$ is technically not a Gramian, but~$E^\T Q E$ is).
In \pyMOR, we can obtain a ROM using BT with
\begin{pythoncode}
    bt = BTReductor(fom)
    rom = bt.reduce(N)
\end{pythoncode}
where the~\PY{reduce} method also allows to prescribe the target~$\Hinf$ error.
\par
Furthermore, there exist different extensions of~BT to structured systems.
Here, we focus on second-order systems \cref{eq:so_fom}.
Writing such systems in first-order form~\eqref{eq:lti_fom} of twice the
size, BT extensions are based on balancing parts of the controllability and
observability Gramians~\cite{MeyS96, ChaLVD06, ReiS08}.
For instance, second-order BT with position balancing (SOBTp)
balances the diagonal blocks of the Gramians corresponding to the position
variable in the first-order form.
Unfortunately, none of the proposed second-order balancing methods provide a
rigorous error bound as~BT in the unstructured case.
\par
SOBTp is available in~\pyMOR via the~\PY{SOBTpReductor}. For a second-order
model~\PY{fom}, we can compute a corresponding~SOBTp~ROM of order~\PY{N}
(resulting in a first-order~ROM of order~$2 \cdot \text{\PY{N}}$) using
\begin{pythoncode}
    sobtp = SOBTpReductor(fom)
    rom = sobtp.reduce(N)
\end{pythoncode}
The resulting~\PY{rom} is again a~\PY{SecondOrderModel} like~\PY{fom}.

\subsection{\texorpdfstring{$\Htwo$}{H2}-optimal model order reduction}

The iterative rational Krylov algorithm (IRKA)~\cite{GugAB08, AntBG10} is a
method for~$\Htwo$-optimal model order reduction of LTI systems \cref{eq:lti_fom}, i.e.,
it tries to find a reduced-order transfer function~$\hat{H}$ that minimizes the
$\Htwo$ error~$\Norm{H - \hat{H}}_{\Htwo}$.
The method is based on the first-order necessary optimality conditions in
interpolation form, which state that~$\hat{H}$ has to be a (bi-tangential)
Hermite interpolant of~$H$ at the poles of~$\hat{H}$ reflected over the
imaginary axis
(and tangential directions determined by the residues of~$\hat{H}$)
in the case of single-input single-output systems (in the general case).
IRKA is then a fixed-point iteration that repeatedly performs interpolation and
chooses the reflected poles as new interpolation points.
Reductors such as the~\PY{IRKAReductor} and its variants,
which are related to the~$\Htwo$ norm, are provided in~\PY{pymor.reductors.h2}.
\par
We mention here the transfer function IRKA (TF-IRKA)~\cite{beattie2012realization},
which extends IRKA from FOMs given by finite-dimensional LTI systems
to any system with a transfer function that is an element of the $\Htwo$ space,
e.g., time-delay systems or PDEs.
To run, TF-IRKA needs to evaluate the transfer function and its derivative at
the reflected poles of the intermediate ROMs,
which can be arbitrary points in the complex plane.
Therefore, TF-IRKA is not a model-based method,
as it does not require access to the underlying model,
but it is also not completely data-driven,
as it cannot work with a fixed amount of data.
It is implemented in \pyMOR within the \PyClass{TFIRKAReductor}.
\par
The idea of a fixed-point iteration has been extended to structured systems.
An IRKA variant for port-Hamiltonian systems \cref{eq:ph_fom} has been proposed~\cite{GugPGBS12},
which preserves the port-Hamiltonian structure and
upon convergence achieves Lagrange interpolation at the reflected poles.
Since it is not based on~$\Htwo$-optimality for structured systems,
it does not necessarily find a locally~$\Htwo$-optimal~ROM.
\par
In~\pyMOR, given a port-Hamiltonian~\PY{fom}, we can perform~PH-IRKA using
\begin{pythoncode}
    phirka = PHIRKAReductor(fom)
    rom = phirka.reduce(N)
\end{pythoncode}
where~\PY{N} is the order of the reduced system.
In order to investigate the performance of the reduced system, the same functionality
as for general~LTI systems (such as~\PY{poles} or~\PY{transfer_function}) is available.
Due to the structure preservation, the~\PY{rom} is a~\PY{PHLTIModel} similar to the~\PY{fom}.

\section{Data-driven MOR methods}\label{sec:data-driven-methods}
In this section we describe the data-driven MOR methods available in~\pyMOR.
In contrast to the model-based methods discussed in the previous section,
these methods construct the~ROM solely from simulation or measurement data generated by the FOM or
underlying physical system.
\par
We begin by introducing proper orthogonal decomposition~(POD) as a general technique for reduced state
approximation from full-order snapshot data (\cref{sec:pod}).
POD is also the basis for the POD-ML approach, which combines POD with machine learning methods for the
approximation of the reduced state (\cref{sec:ml_mor}).
Closely related to POD is Dynamic Mode Decomposition, which identifies reduced dynamical systems from
state space measurements (\cref{sec:dmd}).
Methods based on interpolating the transfer function of LTI control systems are discussed in
\cref{sec:loewner,sec:paaa}.
The eigensystem realization algorithm identifies reduced discrete-time LTI systems from impulse response
measurements (\cref{sec:era}).

\subsection{State approximation via proper orthogonal decomposition}
\label{sec:pod}
Proper orthogonal decomposition (POD)~\cite{Sirovich1987TurbulenceDynamicsCoherent} is a popular method
for constructing reduced state approximation spaces from full-order state space snapshots.
It is applicable to both model-based (\cref{sec:model-based-methods}) and data-driven
(\cref{sec:data-driven-methods}) MOR approaches.
The POD basis for a set of snapshot vectors $x_1, \ldots, x_{n_\textnormal{train}} \in \R^n$ is obtained by collecting
these snapshot vectors in a snapshot matrix~$X \in \R^{n \times n_\textnormal{train}}$, $X_{:,j} = x_j$,
and then computing the first~$N$ left singular vectors $v_1, \ldots, v_N \in \R^n$ of~$X$.
The linear span of these basis vectors is an~$\ell^2$-optimal approximation space for the snapshot data in the sense that
the error
\begin{align*}
    \sum\limits_{j=1}^{n_\textnormal{train}} \min_{v_N \in V_N} \Norm*{x_j - v_N}^2
\end{align*}
is minimal among all possible~$N$-dimensional approximation spaces.
Using~\pyMOR, the POD of a snapshot \VectorArray~\PY{X} can be computed with
\begin{pythoncode}
    V, V_svals = pod(X, modes=N)
\end{pythoncode}
which returns a \VectorArray of the first~\PY{N} singular vectors (POD modes) and a \NumPy array
of associated singular values.
Instead of prescribing a fixed number of modes, different truncation error tolerances based on the
singular value decay of the snapshot matrix~\PY{X} can be prescribed using the~\PY{atol}, \PY{rtol} and
\PY{l2_err} parameters.
The~\PY{product} argument allows passing an inner product \Operator w.r.t.\ which the SVD is computed.
Further, different algorithms (QR decomposition, method of snapshots) for computing the SVD of~\PY{X}
can be selected with the~\PY{method} parameter.
\par
\begin{figure}[htbp]
	\begin{center}
		\resizebox{\columnwidth}{!}{%
			\includegraphics{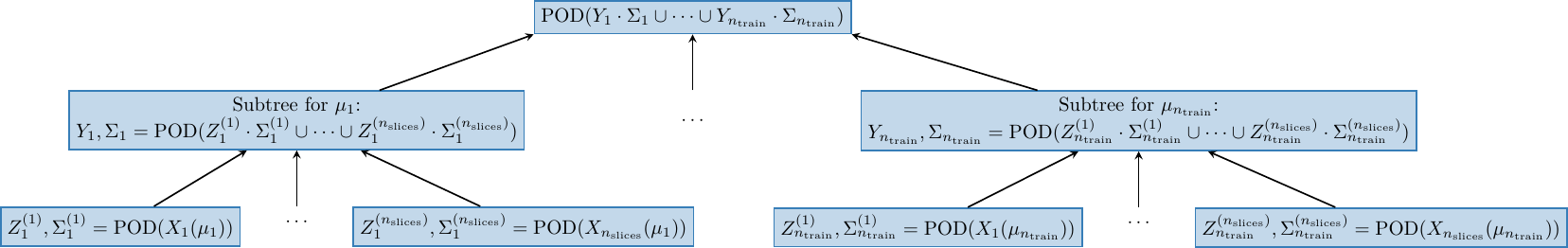}
		}
	\end{center}
	\caption{HAPOD tree for the example in~\Cref{sec:navier_stokes}. The training parameters are denoted as~$\mu_1,\ldots,\mu_{n_\mathrm{train}}\in\mathcal{P}$ with associated solution trajectory cut into slices~$X_1(\mu_i),\ldots,X_{n_\mathrm{slices}}(\mu_i)$ for~$i=1,\ldots,n_\mathrm{train}$. In the leaf nodes, for every such slice the~POD is computed giving POD modes~$Z_i^{(j)}$ for parameter~$\mu_i$ and slice~$j$ together with corresponding singular values~$\Sigma_i^{(j)}$. The POD modes for the different parameters, weighted by the singular values, are then collected and another POD is performed, giving~$Y_i$ as POD modes and~$\Sigma_i$ as singular values for parameter~$\mu_i$. Finally, all weighted modes~$Y_1\cdot\Sigma_1,\ldots Y_{n_\mathrm{train}}\cdot\Sigma_{n_\mathrm{train}}$ are combined for a final POD in the root node of the~HAPOD tree.}
	\label{fig:hapod-tree}
\end{figure}

For large datasets, \pyMOR implements the HAPOD~\cite{HimpeLeibnerEtAl2018HierarchicalApproximateProper}
algorithm, which allows decomposing large POD problems into smaller sub-PODs along arbitrary tree structures.
For instance, for the parametric, time-dependent Navier-Stokes example in~\Cref{sec:navier_stokes}, we use a
computational tree as depicted in~\Cref{fig:hapod-tree}: for each training parameter, we decompose the solution
trajectory into~\PY{hapod_slices} many slices.
The POD modes of these slices, scaled by their singular values, are then the inputs of second-level PODs
for each training parameter. The scaled singular vectors for the individual training trajectories are
then collected for a final third-level POD.
We can use the~\PyClass{Node} class of the~\PY{hapod} module to define this tree as follows:
\begin{pythoncode}
    tree = Node()
    for i_mu in range(len(training_mus)):
        node = tree.add_child(tag=i_mu)
        for i_slice in range(hapod_slices):
            node.add_child(tag=(i_mu, i_slice), after=(i_mu-1,) if i_mu else None)
\end{pythoncode}
Then the POD of the training data can be computed using:
\begin{pythoncode}
    V, V_svals, _ = hapod(tree, get_hapod_slice, std_local_eps(tree, tol, 0.1))
\end{pythoncode}
Here, \PY{get_hapod_slice} is a user function mapping a tree~\PyClass{Node} to the respective input data.
The function can access the node's~\PY{tag} to select the data to be returned.
Since we do not use parallelization in our experiment and want to keep only a single trajectory in memory at a
time, we specify~\PY{after=(i_mu-1,)} to ensure a sequential computation of the PODs.
For parallel computation of the sub-PODs, we could pass any object satisfying the~\PyClass{Executor}
interface from~\PyClass{concurrent.futures} as the~\PY{executor} argument.
The provided~\PY{std_local_eps} function defines local truncation error tolerances as in
\cite{HimpeLeibnerEtAl2018HierarchicalApproximateProper}, ensuring that the mean-squared approximation
error for the computed POD modes stays below~\PY{tol}.

\subsection{Data-driven MOR using POD and machine learning}\label{sec:ml_mor}

A general approach to data-driven MOR of parametric systems is to combine a POD reduced space with machine learning~(ML) of
the map from parameters to reduced solution coefficients:
Starting from state-space trajectories for given training parameters as input, first a POD
approximation space for all snapshot vectors is computed,
and the reduced coefficients of the projection of the training data onto the POD space are determined.
(These coefficients coincide with the right-singular vectors associated to the POD modes.)
In a second step, a supervised~ML algorithm is used to learn the
parameter-to-reduced-coefficients map, where the coefficients of the projected snapshots vectors are used
as training data.
This approach was pioneered in~\cite{hesthaven2018nonintrusive,wang2019nonintrusive}, using artificial
neural networks for the machine learning approximation.
\par
By using a black-box machine learning method, this MOR strategy is universally applicable without requiring an efficient way to compute the reduced coefficients for unseen parameters.
The approach is particularly useful for time-dependent problems, which would require iterative time stepping with a system solve in each iteration.
Moreover, the method is attractive for nonlinear problems, where hyperreduction techniques such as empirical interpolation (\cref{sec:deim}) would be required for model-based approaches.
As access to the underlying code of the high-fidelity solver is limited in many practical applications,
implementing such hyperreduction methods is often infeasible.
\par
Two \Reductors for the~POD-ML approach are available in~\pyMOR.
The~\PyClass{DataDrivenPODReductor} operates on training parameters and a \VectorArray of corresponding
state-space snapshots as input.
The POD of the \VectorArray data is performed using one of~\pyMOR's generic POD algorithms
(\cref{sec:pod}), after which one of the available ML regressors (\cref{sec:regressors}) is used to learn the reduced
coefficients.
The~\PyClass{DataDrivenPODReductor} applied to~\PY{training_parameters} and corresponding~\PY{training_snapshots} can be constructed and used as follows:
\begin{pythoncode}
    regressor = ...  # initialize a regressor, for instance VKOGA, DNN, etc.
    reductor = DataDrivenPODReductor(training_parameters, training_snapshots,
                                     regressor=regressor, pod_params=pod_params)
    rom = reductor.reduce()  # POD of snapshots and ML training

    mu = parameter_space.sample_randomly()  # draw a random test parameter
    U_rom = rom.solve(mu)  # obtain reduced coefficients from the ML surrogate
    U_reconstructed = reductor.lincomb(U_rom)  # reconstruct high-dimensional solution
\end{pythoncode}
For special cases where a POD basis has already been computed or to directly learn the
parameter-to-output map, \PyClass{DataDrivenReductor} is available, which takes \NumPy arrays of reduced
coefficients or outputs as inputs, without any assumptions on interpretability.
In particular, \PyClass{DataDrivenReductor} can be used to build ROM hierarchies where the training data
for the ML ROM is given by reduced solution coefficients of another RB ROM (see \cref{sec:navier_stokes_ml}).
Both \Reductors return a~\PY{DataDrivenModel}, which uses the selected regressor to directly predict the respective quantities.
\par
For time-dependent problems, two different strategies are available (see~\cite{haasdonk2023certified} for a
detailed discussion):
On the one hand, time can be treated as an additional input to the machine learning surrogate, and the output corresponds to the reduced coefficients or output quantity at a particular time instance.
Thus, the ML algorithm learns the map
\begin{align*}
	\Phi\colon\Params\times[0,T]\to\R^N,\qquad (\mu,t)\mapsto\hat{x}(\mu,t).
\end{align*}
We refer to this approach as~``random-access-in-time''.
On the other hand, the whole time trajectory of reduced coefficients can be predicted at once with only the parameter serving as input,
i.e., the map
\begin{align*}
	\Phi\colon\Params\to\R^{N\cdot n_t},\qquad \mu\mapsto[\hat{x}(\mu,t_0),\ldots,\hat{x}(\mu,t_{n_t})]
\end{align*}
is learned.
We call this strategy~``time-vectorized''.
The desired strategy can be selected for both \Reductors by passing the boolean~\PY{time_vectorized} argument.
Both \Reductors support both time-dependent and stationary problems.
Based on the training data provided to the \Reductors, it is automatically deduced whether the problem is
time-dependent or not.
\subsubsection{Supported machine learning algortihms}\label{sec:regressors}
Both \PyClass{DataDrivenReductor} and \PyClass{DataDrivenPODReductor} rely on \PY{regressor} objects that carry out the actual ML algorithm.
Any class fulfilling the \PyClass{Regressor} interface of~\ScikitLearn~\cite{pedregose2011scikit-learn} can be used.
As the default method, we provide a purely~\NumPy-based version of the~vectorial kernel orthogonal greedy
algorithm~(VKOGA) in~\pyMOR (see~\cite{santin2021kernel} for a general overview on kernel methods and~VKOGA).
The implementation utilizes~\pyMOR's weak greedy algorithm for iterative selection of centers in the kernel expansion based on the~$f$-, $f\cdot P$- or~$P$-greedy selection criteria~\cite{wenzel2022analysis}.
Therefore, it is not required to install any additional machine learning-related library in order to use the~\PY{VKOGARegressor}.
In addition, there is a builtin implementation of deep neural networks~(DNNs) based on~\PyTorch~\cite{paszke2019pytorch} with customizable neural network architecture and training algorithm.
The~\ScikitLearn library provides several further regressors that can be directly used without any adjustments.
For instance, we employ Gaussian process regression~(GPR)~\cite{rasmussen2005gaussian} for the numerical experiment in~\Cref{sec:navier_stokes_ml}.
\par
When calling the~\PY{reduce}-method of the \Reductor, the training data is prepared and the~\PY{fit}-method of the regressor is executed on the training data.
The resulting \PyClasses{DataDrivenModel} use the regressor's \PY{predict}-method for solution or output prediction.
As any other \ScikitLearn regressor, \PY{VKOGARegressor} can also be used directly for general ML
applications:
\begin{pythoncode}
    regressor = VKOGARegressor(kernel=GaussianKernel(length_scale=0.3),
                               criterion='fp', max_centers=200, tol=1e-6, reg=1e-12)
    regressor.fit(X, F)
\end{pythoncode}
Here,~\PY{X} and~\PY{F} are given (vector-valued) input-output pairs.
Instead of using the Gaussian kernel provided by~\pyMOR, it is possible to pass any kernel function available in the \ScikitLearn module \PY{sklearn.gaussian_process.kernels} via the~\PY{kernel} argument.
In addition, it is possible to pass~\PY{'f'}, \PY{'fp'} or~\PY{'p'} as~\PY{criterion} in order to change the greedy selection criterion.
As termination criteria for the greedy search, the maximum number of centers can be restricted
via~\PY{max_centers} or an error tolerance can be prescribed using the~\PY{tol}-argument.
The regularization parameter of the kernel interpolation can be adjusted by means of the~\PY{reg}-parameter.
For further analysis, additional information can be extracted from the kernel surrogate, such as the center locations via~\PY{regressor._surrogate._centers} or the indices of the centers in the training set via~\PY{regressor._surrogate._centers_idx}.
\par
The~\PY{NeuralNetworkRegressor} makes use of~\PyTorch and implements a training algorithm with several standard features for neural network training, such as multiple restarts with different initializations of weights and biases, random splitting of the training data into training and validation sets or early stopping of the optimization based on the validation loss.
It is further possible to customize the builtin neural network in terms of its number of layers, neurons per layer and activation function or to employ a neural network constructed specifically for the task at hand.
There are several additional training parameters such as the optimizer to use, the number of epochs, the
learning rate, the loss function, the number of restarts, the weight decay or a learning rate scheduler that can be adjusted and tuned.
The respective functions and objects provided by~\PyTorch can be used directly to replace the default settings.
\par
Besides the possibility to choose between different ML algorithms, it is also possible to scale the inputs or outputs by using scaling algorithms from~\ScikitLearn{}'s~\PY{preprocessing}-module
(see~\Cref{sec:navier_stokes_ml} for an example).
It is further possible to iteratively extend the training data by adding more parameters and corresponding solutions or outputs.
Calling the~\PY{reduce}-method of the \Reductor again then trains an updated regressor.
In~\Cref{sec:navier_stokes_ml}, this feature will be used to analyze the performance of the resulting~ROMs in dependence on the amount of given training data.

\subsection{Dynamic mode decomposition}\label{sec:dmd}
Dynamic mode decomposition (DMD) is an identification method for (reduced) linear dynamical systems.
Given a sequence of snapshots~$x^{(0)},\ldots,x^{(n_t)}\in\R^n$ with a fixed time step size~$\Delta t>0$, the idea of~DMD is to compute a matrix~$A\in\R^{n\times n}$ describing a discrete-time dynamical system satisfying
\begin{align*}
	x^{(k+1)} \approx Ax^{(k)}\qquad\text{for }k=0,\ldots,n_t-1.
\end{align*}
Denoting by~$X_1\coloneqq [x^{(0)},x^{(1)},\ldots,x^{(n_t-1)}]\in\R^{n\times n_t}$, $X_2\coloneqq [x^{(1)},\ldots,x^{(n_t)}]\in\R^{n\times n_t}$ the (shifted) data matrices, we can define~$A$ as~$X_2X_1^+$,
where~$X_1^+$ denotes the Moore-Penrose pseudo-inverse of~$X_1$.
In practice, the matrix~$A$ is never constructed explicitly, but a reduced operator~$\hat{A}\in\R^{N\times N}$ with~$N\ll n$ is computed using a truncated~SVD of the matrix~$X_1$.
To be more precise, we compute the truncated~SVD of~$X_1$ as~$X_1\approx U\Sigma V^\T$ with orthonormal matrices~$U\in\R^{n\times N}$ and~$V\in\R^{n_t\times N}$ and a diagonal matrix~$\Sigma\in\R^{N\times N}$.
We then derive the reduced system matrix as~$\hat{A}=U^\T X_2V\Sigma^{-1}\in\R^{N\times N}$.
Let us denote the eigenvalues and eigenvectors of~$\hat{A}$ by~$\lambda_1,\dots,\lambda_N\in\C$ and~$y_1,\ldots,y_N\in\C^N$, i.e., $\hat{A}y_j=\lambda_j y_j$ for~$j=1,\ldots,N$.
The corresponding~DMD~modes can then be reconstructed as~$w_j=X_2V\Sigma^{-1}y_j\in\C^n$.
This definition of~$w_1,\ldots,w_N$ corresponds to the so-called exact~DMD.
We sort the~DMD~eigenvalues by phase, i.e.~by the absolute value of the angle of the~DMD~eigenvalues from the positive real axis on the complex plane.
\par
The reconstructed solution of the reduced dynamical system yields an approximation to the continuous-time dynamics for~$t\in[0,T]$ as
\begin{align}\label{equ:dmd-approximation}
	x(t) \approx \sum\limits_{j=1}^{N}b_je^{\omega_j t}w_j,
\end{align}
where~$\omega_j=\log(\lambda_j)/{\Delta t}\in\C$ for~$j=1,\ldots,N$ are the continuous-time~DMD eigenvalues.
The coefficient vector~$b\in\R^N$ is given as the coefficients of the projection of the initial condition~$x^{(0)}\in\R^n$ onto the span of the~DMD modes, i.e.~$b=(W^*W)^{-1}W^*x^{(0)}\in\C^N$, where~$W=[w_1,\ldots,w_N]\in\C^{n\times N}$.
The~DMD~modes are typically not orthonormal and $W^*W\in\R^{N\times N}$ might be ill-conditioned.
Thus, we compute~$b$ in practice by solving the least squares problem~$\min_{b\in\C^N}\,\lVert Wb-x^{(0)}\rVert_2^2$ to ensure stability.
\par
The~DMD implementation in~\pyMOR makes use of one of~\pyMOR's~\PY{svd} methods for \VectorArrays: the user can choose between an~SVD computed via QR decomposition or the method of snapshots~\cite{Sirovich1987TurbulenceDynamicsCoherent}.
It is possible to either specify a truncation tolerance or a fixed truncation rank for the~SVD of~$X_1$.
In order to obtain the coefficients~$b$, we use~\pyMOR's~\PY{QRLeastSquaresSolver}.
The following code snippet performs~DMD on a~\PY{VectorArray}~\PY{X} and computes a~DMD approximation according to~\Cref{equ:dmd-approximation},
similar to how DMD is used for the example in~\Cref{sec:navier_stokes_dmd}:

\begin{pythoncode}
    W, omegas = dmd(X, modes=num_dmd_modes, order='phase', cont_time_dt=fom.dt)

    # compute reduced coefficients for initial data
    lstsq_solver = QRLeastSquaresSolver()
    b = lstsq_solver.solve(VectorArrayOperator(W), X[0])

    # compute continuous-time dynamics
    t = np.arange(len(X)) * fom.dt
    exp_factors = np.exp(np.outer(omegas, t))
    coeffs = b.to_numpy() * exp_factors
    X_dmd = W.lincomb(coeffs)
\end{pythoncode}
In this example, we use a fixed truncation rank by passing the~\PY{modes}-argument in the~\PY{dmd} call.
Moreover, we order the DMD eigenvalues according to their phase before truncation.
By setting~\PY{cont_time_dt} to~\PY{fom.dt} we ensure that continuous-time DMD eigenvalues~$\omega_j$ are returned that include the logarithmic transformation of the eigenvalues~$\lambda_j$ of~$\hat{A}$ and the scaling by~$\Delta t$, the time step size of the corresponding~\PY{fom}.

\subsection{Loewner framework}\label{sec:loewner}

The Loewner framework~\cite{MayA07} is a family of methods for constructing
LTI systems from transfer function data (whether from a model or measurements),
including extensions to parametric~\cite{IonA14} and certain classes of
nonlinear systems~\cite{AntGI16,GosA18}.
It is based on Loewner matrices to directly construct a model that interpolates
the data, with a possible truncation in the case of redundant data.
\par
To illustrate the method, let us focus on the single-input single-output case
and suppose we are given transfer function data
$(s_i, H_i)$, where $s_i, H_i \in \C$, for $i = 1, 2, \ldots, 2 d$,
and the goal is to construct a ROM with a transfer function $\hat{H}$ such that
$\hat{H}(s_i) \approx H_i$.
The first step is to split the data into two parts
(for simplicity of equal sizes)
$\{(\lambda_1, F_1), \ldots, (\lambda_d, F_d)\}$ and
$\{(\mu_1, G_1), \ldots, (\mu_d, G_d)\}$
such that~$\{(\lambda_1, F_1), \ldots, (\lambda_d, F_d)\}
\cup \{(\mu_1, G_1), \ldots, (\mu_d, G_d)\}
= \{(s_1, H_1), \ldots, (s_{2 d}, H_{2 d})\}$.
Next, we construct the Loewner and the shifted Loewner matrices
\begin{equation*}
    \mathbb{L} =
    \begin{bmatrix}
        \frac{G_1 - F_1}{\mu_1 - \lambda_1}
        & \cdots
        & \frac{G_1 - F_d}{\mu_1 - \lambda_d} \\
        \vdots
        & \ddots
        & \vdots \\
        \frac{G_d - F_1}{\mu_d - \lambda_1}
        & \cdots
        & \frac{G_d - F_d}{\mu_d - \lambda_d}
    \end{bmatrix}
    \quad \text{and} \quad
    \sigma\mathbb{L} =
    \begin{bmatrix}
        \frac{\mu_1 G_1 - \lambda_1 F_1}{\mu_1 - \lambda_1}
        & \cdots
        & \frac{\mu_1 G_1 - \lambda_d F_d}{\mu_1 - \lambda_d} \\
        \vdots
        & \ddots
        & \vdots \\
        \frac{\mu_d G_d - \lambda_1 F_1}{\mu_d - \lambda_1}
        & \cdots
        & \frac{\mu_d G_d - \lambda_d F_d}{\mu_d - \lambda_d}
    \end{bmatrix}.
\end{equation*}
Note that the Loewner matrix $\mathbb{L}$ is a divided differences matrix for
the unknown transfer function~$H(s)$,
while the shifted Loewner matrix $\sigma\mathbb{L}$ is a divided difference
matrix for the ``shifted'' transfer function $s H(s)$
(in the discrete-time case, multiplication by $s$ in the frequency domain does
correspond to the shift in the time domain).
Then, the LTI system with
\begin{equation*}
    \hat{E} = -\mathbb{L}, \quad
    \hat{A} = -\sigma\mathbb{L}, \quad
    \hat{B} =
    \begin{bmatrix}
        G_1 \\
        \vdots \\
        G_d \\
    \end{bmatrix}, \quad
    \hat{C} =
    \begin{bmatrix}
        F_1 & \cdots & F_d
    \end{bmatrix}.
\end{equation*}
achieves interpolation $\hat{H}(s_i) = H_i$ and
$\hat{H}$ is of minimal possible degree
if $s_i \mathbb{L} - \sigma\mathbb{L}$ is of full rank for all $s_i$,
Typically, Loewner matrices have rapidly decaying singular values,
and an SVD of $\mathbb{L}$ and/or~$\sigma\mathbb{L}$ is used to truncate the
above model, which then only achieves approximate interpolation.
\par
There are many generalizations and extensions of the above procedure,
e.g., for Hermite interpolation.
Currently, \pyMOR provides a \PyClass{LoewnerReductor} implementing Lagrange
interpolation for multiple-input multiple-output (MIMO) systems,
with different options for data splitting (automatic or user-specified) and
handling MIMO systems (matrix or tangential interpolation).

\subsection{Parametric adaptive Antoulas-Anderson method}\label{sec:paaa}

Similarly as the Loewner framework discussed above,
the adaptive Antoulas-Anderson (AAA) method~\cite{NakST18}
constructs a rational approximation from data in the complex domain.
It uses a greedy approach to select which data to interpolate and
linearized least-squares to fit the other data.
\par
Briefly,
it is based on the barycentric form of rational functions for interpolation,
specifically, given a subset of the data $\{(s_1, H_1), \ldots, (s_d, H_d)\}$,
the rational function
\begin{equation*}
    \hat{H}(s) =
    \frac{
        \sum_{i = 1}^{d} \frac{\alpha_i H_i}{s - s_i}
    }{
        \sum_{i = 1}^{d} \frac{\alpha_i}{s - s_i}
    }
\end{equation*}
achieves interpolation $\hat{H}(s_i) = H_i$ for any choice of scalars
$\alpha_i \neq 0$.
The free parameters $\alpha_i$ can be used to interpolate at additional $d$ data
points, which can be formulated using Loewner matrices.
Alternatively, as done in AAA, the parameters $\alpha_i$ can be selected to
reduce the least-squares error at other data points.
On top of that, AAA selects interpolation points one by one in a greedy fashion,
trying to minimize the $L_{\infty}$ error in the data.
\par
The~AAA method has been extended in many directions.
An extension to parametric systems, called p-AAA~\cite{RodBG23},
has been implemented in \pyMOR.
It uses the multivariate barycentric form that similarly combines interpolation
and least-squares fitting to build an approximate parametric transfer function.
The \PyClass{PAAAReductor} in \pyMOR can be used to apply both AAA and p-AAA for MIMO
systems.

\subsection{Eigensystem realization algorithm}\label{sec:era}
The eigensystem realization algorithm (ERA)~\cite{juang1985eigensystem}
is a system identification method for discrete-time LTI systems,
where the state-space realization is recovered form the impulse response.
In more detail, for a discrete-time LTI system
\begin{equation*}
    \begin{aligned}
        x^{(k + 1)} & = A x^{(k)} + B u^{(k)}, & k \ge 0, \\
        x^{(0)} & = 0, \\
        y^{(k)} & = C x^{(k)} + D u^{(k)}, & k \ge 0,
    \end{aligned}
\end{equation*}
its impulse response is given by
\begin{equation*}
    h(k) =
    \begin{cases}
        C A^{k - 1} B, & k > 0, \\
        D, & k = 0, \\
        0, & k < 0.
    \end{cases}
\end{equation*}
Then, from the data $h(0) = D, h(1) = C B, \ldots, h(d) = C A^{d - 1} B$,
using Hankel matrices and SVD,
ERA constructs matrices $A, B, C, D$.
\par
For a continuous-time LTI system~\eqref{eq:lti_fom} with $E = I$,
we suppose that the data is the matrix $D$ and samples of the impulse response
$C B, C e^{\Delta t A} B, \ldots, C e^{d \Delta t A} B$,
where $\Delta t > 0$ is a sampling time
(we additionally scale the impulse response by $\Delta t$ to correspond to the
impulse response of the discrete-time system,
but it does not change basic idea).
Using ERA to construct $e^{\Delta t A}$, $B$, and $C$,
we use the matrix logarithm to recover $A$.
\par
In \pyMOR, the \PyClass{ERAReductor} implements ERA for MIMO systems with
support for tangential projection for more efficient handling of systems with
many inputs or outputs~\cite{KraG16}.

\section{Numerical examples}\label{sec:experiments}

In this section we discuss two test cases.
We start with a parametrized Navier-Stokes model, for which we first consider a model-based RB approach
(\cref{sec:navier_stokes_pod_deim}) that we compare with the data-driven POD-ML method
(\cref{sec:navier_stokes_ml}).
We also consider a combination of both approaches, where the model-based ROM is used for efficient
training data generation for the POD-ML ROM.
Further, we explore~DMD to approximate time-periodic vortex shedding patterns (\cref{sec:navier_stokes_dmd}).
\par
In the second numerical experiment, we focus on system-theoretic~MOR methods applied to a mass-spring-damper chain control system.
The system under consideration can be formulated as a second-order system as well as a port-Hamiltonian
system and we compare classical model-based methods such as~BT and~IRKA to their structure-preserving variants.
We compare these methods to the data-driven approaches implemented in \pyMOR, i.e.~the Loewner framework,
AAA and~ERA (\cref{sec:msd-nonparametric}).
By considering a parametric version of the control system, we also investigate the parametric versions of BT and IRKA and the~p-AAA algorithm in practice
(\cref{sec:msd-parametric}).
\par
All experiments were executed as single-threaded processes on a workstation with Intel Xeon~Gold~6254
CPUs running Ubuntu~22.04. The maximum amount of required RAM per experiment was 52GB.

\subsection{Incompressible Navier-Stokes equations}\label{sec:navier_stokes}
\begin{figure}[htbp]
	\centering
	\includegraphics{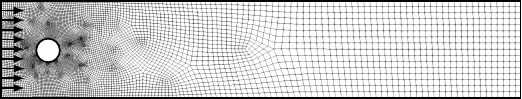}
	\caption{Navier-Stokes example: Computational domain $\Omega$ and finite element grid.}
	\label{fig:navier-stokes-domain}
\end{figure}

\begin{figure}
    \centering
    \includegraphics{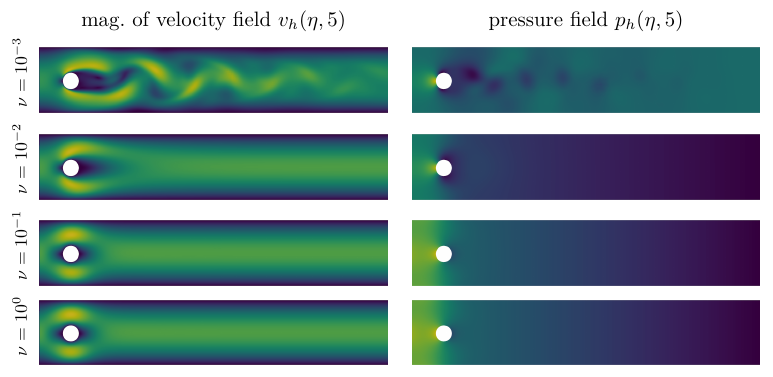}
    \caption{Navier-Stokes example: FOM simulations for different kinematic viscosities~$\nu$. Visualized
    are the magnitude of velocity field~$v_h$ and pressure field~$p_h$ at time~$t=5$.}
    \label{fig:navier_stokes_solutions}
\end{figure}

We consider the incompressible Navier-Stokes equations
\begin{equation}
    \label{eq:navier_stokes}
    \begin{aligned}
        \partial_t v(\eta, t) - \nu \Delta_\eta v(\eta, t)
            + v(\eta,t)\nabla_\eta v(\eta,t)
            + \nabla_\eta p(\eta) &= 0  && \eta \in \Omega, t \in [0,T],\\
        \nabla_\eta\cdot v(\eta,t)  &= 0 && \eta \in \Omega, t \in [0,T],
    \end{aligned}
\end{equation}
with~$T=8$ and~$\Omega$ the ``flow around a cylinder'' benchmark geometry from
\cite{SchaferTurekEtAl1996BenchmarkComputationsLaminar}, given by
\begin{align*}
    \Omega = (0, 2.2) \times (0, 0.41) \setminus D, \qquad D\coloneqq \set*{\eta \in \R^2 \given \lvert\eta - (0.2,0.2)\rvert^2 \leq 0.05^2},
\end{align*}
and depicted in~\Cref{fig:navier-stokes-domain}.
As in~\cite{SchaferTurekEtAl1996BenchmarkComputationsLaminar}, we prescribe non-slip boundary conditions~$v(\eta,t) = 0$ on the lower and upper
walls ($\eta_2 = 0$ and~$\eta_2 = 0.41$) and the obstacle boundary~$\partial D$.
On the left boundary of the domain~$(\eta_1 = 0)$, we prescribe a time-dependent Poiseuille inflow velocity profile
of the form
\begin{equation}
    \label{eq:inflow_condition}
    v\left(\begin{bmatrix}0 & \eta_2\end{bmatrix}^\T, t\right) = \sin(\pi\cdot t/8) \cdot 1.5 \cdot 4 \cdot \frac{\eta_2\cdot(0.41-\eta_2)}{(0.41)^2}.
\end{equation}
As initial condition, we let~$v(\eta, 0) = 0$.
For the parametric MOR experiments, we consider the kinematic velocity~$\mu \coloneqq  \nu$ as parameter of
interest. To avoid challenges due to slow Kolmogorov~$N$-width decay
\cite{OhlbergerRave2016ReducedBasisMethods}, we restrict ourselves to a
parameter domain~$\nu \in \mathcal{P} \coloneqq  [10^{-2}, 10^0]$ of relatively high viscosity.
For the DMD experiment (\Cref{sec:navier_stokes_dmd}), we consider a viscosity of~$\nu = 10^{-3}$, for
which vortex shedding occurs (top row of~\Cref{fig:navier_stokes_solutions}).
\par
\begin{figure}
    \begin{center}
        \includegraphics{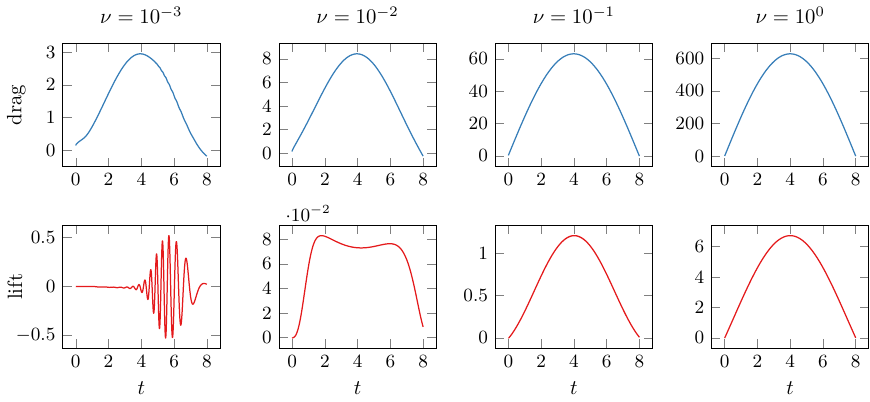}
    \end{center}
    \caption{Navier-Stokes example: Drag and lift vs.\ time of~FOM simulations for different viscosities~$\nu$.}
    \label{fig:navier_stokes_outputs}
\end{figure}

We use a Taylor-Hood finite element discretization implemented with \FEniCSx.
The mesh with~7,808 quadrilateral elements leads to a discrete state vector~$x(t)\coloneqq \begin{bmatrix}v_h(\cdot,t) & p_h(\cdot,t)\end{bmatrix}^\T$ with~71,352 DoFs (see~\Cref{fig:navier-stokes-domain}).
We use an IMEX time discretization (\cref{sec:imex_models}) with an equidistant time-step size of~$\Delta t = 0.0005$, resulting in~$n_t \coloneqq  T / \Delta t = 16,000$ time steps.
Denoting by~$\varphi_j$ the finite element basis of the velocity space and by~$\psi_j$ the finite element
basis of the pressure space, the matrix~$E$ is given by
\begin{align*}
    \begin{bmatrix}
        E^{(vv)} & 0 \\ 0 & 0
    \end{bmatrix}
    \quad\text{where}\quad
    E^{(vv)}_{j,i} = \int_\Omega \varphi_i(\eta)\varphi_j(\eta) \deta.
\end{align*}
The implicit operator corresponding to the Stokes terms in~\Cref{eq:navier_stokes} is given by
\begin{align*}
    F_i(\nu) = \begin{bmatrix}\nu \cdot F_i^{(vv)} & \left(F_i^{(pv)}\right)^\T \\ F_i^{(pv)} & 0\end{bmatrix}
\end{align*}
with
\begin{align*}
    \begin{aligned}
        \left[F_i^{(vv)}\right]_{j,i} = \int_{\Omega} -\nabla \varphi_i(\eta)\cdot \nabla \varphi_j(\eta) \deta
        \quad\text{and}\quad
        \left[F_i^{(pv)}\right]_{j,i} = \int_\Omega \nabla \cdot \varphi_i(\eta) \psi_j(\eta) \deta.
    \end{aligned}
\end{align*}
The explicit operator
$f_e = \begin{bmatrix}f_e^{(v)} & f_e^{(p)}\end{bmatrix}^\T$
corresponding to the advection term in~\Cref{eq:navier_stokes} and the inflow
boundary condition~\Cref{eq:inflow_condition} is given by
\begin{align*}
    \begin{aligned}
        \left[f_e^{(v)}\left(t_k, \begin{bmatrix}v_h & p_h\end{bmatrix}^\T, \nu\right)\right]_j
            &=  \left[- \frac{1}{\Delta t} E^{(vv)}\cdot (v^{(n+1)}_{h,\text{in}} - v^{(k)}_{h,\text{in}})
                        + \nu\cdot F_i^{(vv)} \cdot v^{(n+1)}_{h,\text{in}} \right]_j \\
        &  \qquad\qquad - \int_\Omega (v_h(\eta) + v^{(k)}_{h,\text{in}}(\eta)) \nabla (v_h(\eta) + v^{(k)}_{h,\text{in}}(\eta))
                            \cdot \varphi_j(\eta) \deta, \\
        \left[f_e^{(p)}\left(t_k, \begin{bmatrix}v_h & p_h\end{bmatrix}^\T, \nu\right)\right]_j
            &= \left[F_i^{(pv)} \cdot v^{(n+1)}_{h,\text{in}}\right]_j.
    \end{aligned}
\end{align*}
Here, we have split the velocity part~$v_h^{(k)}$ of the solution as
\begin{align*}
    v_h^{(k)} = v^{(k)}_{h,\text{hom}} + v^{(k)}_{h,\text{in}},
\end{align*}
where~$v_{h,\text{in}}^{(k)}$ is a lifting of the boundary condition~\Cref{eq:inflow_condition}
at time~$t_k$ to a finite element function on~$\Omega$ such that~$v^{(k)}_{h,\text{hom}}$ vanishes on the inflow boundary.
The~FOM is then given by~\Cref{eq:imex}, where the solution state vector~$x^{(k)}$ is of the form
\begin{align*}
    x^{(k)} = \begin{bmatrix}v^{(k)}_{h,\text{hom}} &  p^{(k)}_h\end{bmatrix} ^\T.
\end{align*}
State-space solutions of the~FOM for different values of~$\nu$ are depicted in
\Cref{fig:navier_stokes_solutions}.
As outputs~$g = \begin{bmatrix}g_{\text{drag}} & g_{\text{lift}} \end{bmatrix}$,
we consider the drag and lift coefficients of the obstacle~$D$ given by
\begin{align*}
    \begin{aligned}
        g_{\text{drag}}\left(\begin{bmatrix}v_h & p_h\end{bmatrix}^\T, \nu\right) &=
            \frac{2}{0.1}\cdot\int_{\partial D}
                \hphantom{-}\nu\cdot \nabla v_{h,\parallel}(s) \cdot n(s) n_2(s)
                - p_h(s) n_1(s) \ds,\\
        g_{\text{lift}}\left(\begin{bmatrix}v_h & p_h\end{bmatrix}^\T, \nu\right) &=
            \frac{2}{0.1}\cdot\int_{\partial D}
                -\nu\cdot \nabla v_{h,\parallel}(s) \cdot n(s) n_1(s)
                - p_h(s) n_2(s) \ds,
    \end{aligned}
\end{align*}
where~$n(s)$ denotes the unit outer normal to~$\partial D$ at~$s$ and~$v_{h,\parallel}(s)$ the tangential
component of~$v_h$ at~$s$.
The drag and lift coefficients for~FOM solutions with varying viscosity~$\nu$ are shown in
\Cref{fig:navier_stokes_outputs}.

\subsubsection{Error measures}\label{sec:navier_stokes_error_measures}
For the experiments in the following subsections, we consider the relative~$\ell^2$-in time MOR error
\begin{align}
    \NSRelErrState &\coloneqq
        \frac{\left(\sum_{k=1}^{n_t}
                    \Norm*{x^{(k)}(\nu) - \hat{x}^{(k)}(\nu)}^2
              \right)^{1/2}}
              {\left(\sum_{k=1}^{n_t}
                    \Norm*{x^{(k)}(\nu)}^2
              \right)^{1/2}}, \label{eq:nsrelerrstate}\\
    \intertext{with $\hat{x}$ the respective reduced state. Here, $\Norm*{\cdot}$ is the Euclidean norm on the discrete solution state space.
    Similarly, we define the relative output error measures}
    \NSRelErrDrag &\coloneqq
        \frac{\left(\sum_{k=1}^{n_t}
                    \left\lvert g_{\text{drag}}(x^{(k)}(\nu), \nu) - g_{\text{drag}}(\hat{x}^{(k)}(\nu), \nu)\right\rvert^2
              \right)^{1/2}}
              {\left(\sum_{k=1}^{n_t}
                    \left\lvert g_{\text{drag}}(x^{(k)}(\nu), \nu)\right\rvert^2
              \right)^{1/2}}, \label{eq:nsrelerrdrag}\\
    \NSRelErrLift &\coloneqq
        \frac{\left(\sum_{k=1}^{n_t}
                    \left\lvert g_{\text{lift}}(x^{(k)}(\nu), \nu) - g_{\text{lift}}(\hat{x}^{(k)}(\nu), \nu)\right\rvert^2
              \right)^{1/2}}
              {\left(\sum_{k=1}^{n_t}
                    \left\lvert g_{\text{lift}}(x^{(k)}(\nu), \nu)\right\rvert^2
              \right)^{1/2}}.\label{eq:nsrelerrlift}
\end{align}
We will evaluate these quantities over a test set~$\mathcal{S}_{\text{test}} \subset \mathcal{P}$ of 20
log-uniformly sampled test parameters~$\nu$.
In particular, we will consider the maximum errors over this test set given by
\begin{equation}
    \label{eq:nsrelerrmax}
    \NSRelErrMaxState \coloneqq
        \max_{\nu \in \mathcal{S}_{\text{test}}} \NSRelErrState(\nu),
    \qquad
    \NSRelErrMaxDrag \coloneqq
        \max_{\nu \in \mathcal{S}_{\text{test}}} \NSRelErrDrag(\nu),
    \qquad
    \NSRelErrMaxLift \coloneqq
        \max_{\nu \in \mathcal{S}_{\text{test}}} \NSRelErrLift(\nu).
\end{equation}

\subsubsection{POD-DEIM ROM}\label{sec:navier_stokes_pod_deim}
We first consider a model-based RB approach using least-squares residual minimization
(\Cref{sec:imex_lspg}).
The reduced basis is determined using POD (\Cref{sec:pod}), and we use DEIM for hyperreduction (\Cref{sec:deim}).
\par

\begin{figure}
    \begin{center}
        \includegraphics{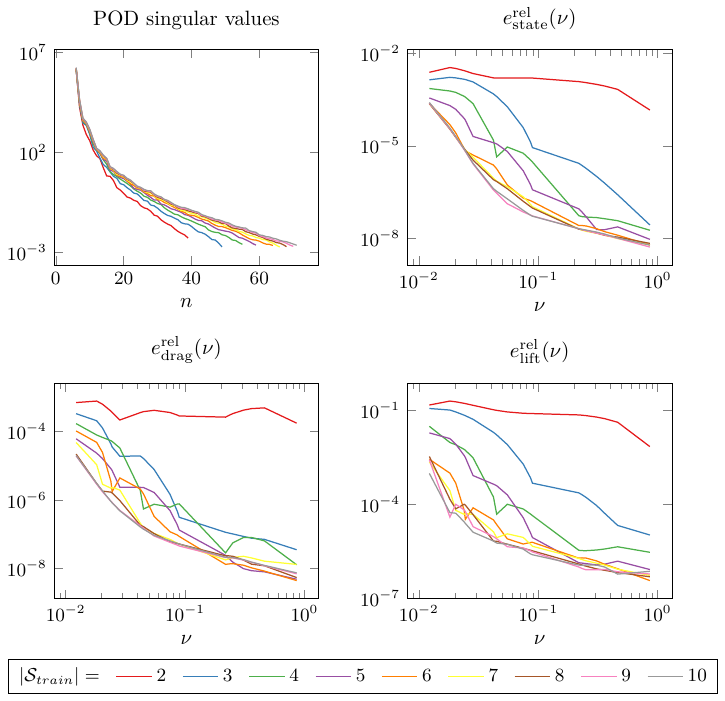}
    \end{center}
    \caption{Navier-Stokes example: POD singular values computed by HAPOD (top left) and
    relative POD-DEIM errors for solution state, drag and lift given by~\Cref{eq:nsrelerrstate,eq:nsrelerrdrag,eq:nsrelerrlift} (other plots).}
    \label{fig:pod_deim_svals_rel_errs}
\end{figure}

Since we use an \PyClass{IMEXModel} for the FOM (\cref{sec:imex_models}) instead of one of~\pyMOR's built-in
\Model classes, we need to specify how to project each of the
Model's \Operators according to~\Cref{sec:imex_lspg} in a custom~\PyClass{Reductor}:
\begin{pythoncode}
class LSPGDeimReductor(BasicObject):  # slightly simplified
    def __init__(self, fom, W, V, Z, I):
        self.__auto_init(locals())

    def reduce(self):
        fom, W, V, Z, I = self.fom, self.W, self.V, self.Z, self.I

        f_e_tw     = EmpiricalInterpolatedOperator(fom.f_e, I, Z, triangular=False)
        E_hat      = project(fom.E,    W,    V)
        F_i_hat    = project(fom.F_i,  W,    V)
        f_e_tw_hat = project(f_e_tw,   W,    V)
        g_hat      = project(fom.g,    None, V)

        return IMEXModel(E_hat, F_i_hat, f_e_tw_hat, g_hat, T=fom.T, dt=fom.dt,
                         solver=ScipyQRLSTSQSolver())
\end{pythoncode}

\PyClass{LSPGDeimReductor} can reduce any~\PyClass{IMEXModel} given to it.
In our case, the FOM is implemented using \FEniCSx.
We expose the \Library{UFL} forms constituting the FOM as \pyMOR \Operators using~\pyMOR's included \FEniCSx bindings:
\begin{pythoncode}
class FenicsxNavierStokesModel(IMEXModel):
    def __init__(self):
        ...  # FEniCSx code to create mesh and define UFL forms

        E = FenicsxMatrixBasedOperator(mass, None, bcs).assemble()
        F_i = (FenicsxMatrixBasedOperator(-stokes_diff, None, bcs).assemble()
                    * ProjectionParameterFunctional('nu', 1)
               + FenicsxMatrixBasedOperator(-stokes_offdiag, None, bcs).assemble())

        class ExplicitOperator(FenicsxOperator):
            def _restrict_form(self, submesh):
                V, V0, _, V_u, _ = spaces(submesh)
                bcu_inflow, U_inlet = inlet(V, V0, V_u)
                rhs, U_, mu, t = rhs_form(submesh, V, U_inlet)
                return rhs, U_, {'nu': nu, 't': t}
        f_e = ExplicitOperator(rhs, U_, {'nu': nu, 't': t}, bcs=bcs, alpha=0.)

        ...  # wrap output forms
        solver = FenicsxLinearSolver(mesh.comm, method='preonly', preconditioner='lu')
        super().__init__(E, F_i, f_e, g, T, dt, solver,
                         visualizer=FenicsxNavierStokesVisualizer(V, dt))
\end{pythoncode}
Here, we use \PyClass{FenicsxMatrixBasedOperator} to wrap the linear \Library{UFL} forms (\PY{mass},
\PY{stokes_diff},\\\PY{stokes_offdiag}) and \PyClass{FenicsxOperator} for the nonlinear form (\PY{rhs}).
In the case of \PY{F_i}, we make \pyMOR aware of its parameter-separable structure
\Cref{eq:parameter_separable} by writing it as a linear combination of \Operators and
\PyClasses{ParameterFunctional}.
To properly support the \PY{restricted} evaluation of \PY{f_e}, we override the \PY{_restrict_form}
method of \PyClass{FenicsxOperator} to recreate the respective~\Library{ufl} form over the submesh
corresponding to the selected interpolation points.
As linear solver, we choose the default direct sparse solver implemented by \Library{PETSc}, which is the
linear algebra backend used by \FEniCSx.
\par
After these preparations, we can now realize our MOR scheme.
First, we compute snapshot data and build a POD basis~$V$ from it.
Since storing a single solution trajectory already requires more than 8GB of memory, we follow the HAPOD
approach described in~\Cref{sec:pod}.
For our experiment, we use a varying number of log-uniformly distributed training parameters~$\nu$.
We prescribe a mean-$\ell^2$ error of~$10^{-4}$, resulting in a reduced space~$V_N$ with dimension
ranging from~$34$ (2 snapshots) to~$66$
(10 snapshots), see~\Cref{tab:pod_deim}.
The singular value decay is depicted in~\Cref{fig:pod_deim_svals_rel_errs}.
\par
Given~$V$, we next compute the range basis~$W$ as discussed in~\Cref{sec:imex_lspg}.
We use~$Z\coloneqq W$ as interpolation basis for~$f_e$ and determine the corresponding interpolation DoFs
using the Q-DEIM algorithm:
\begin{pythoncode}
    W       = estimate_image([fom.E, fom.F_i], [], V)
    I, _, _ = qdeim(W, pod=False)
\end{pythoncode}
The resulting number of interpolation DoFs~$M$, which equals the number of columns of~$W$,
is listed in~\Cref{tab:pod_deim}.
Since~$F_i$ is parameter-separable and can be written as a linear combination of two non-parametric
operators, we always have~$M = N + 2N = 3N$.
\par
After having computed all reduction data, we obtain the~ROM using the previously defined reductor:
\begin{pythoncode}
    reductor = PodDeimReductor(fom, W, V, W, I)
    rom = reductor.reduce()
\end{pythoncode}
We document the MOR errors~\Cref{eq:nsrelerrmax} in~\Cref{tab:pod_deim}.
We observe that the MOR error decay stagnates at around 6 training snapshots with relative errors of
order~$10^{-4}$ for the solution and drag, as well as~$10^{-3}$ for the lift.
This suggests that 6 training snapshots is sufficient to represent the entire solution manifold~$\set{x^{(k)}(\nu) \given 1\leq
n \leq n_t, \nu \in \mathcal{P}}$ for the prescribed error tolerance.
\Cref{fig:pod_deim_svals_rel_errs} shows that all error measures increase as the viscosity~$\nu$ decreases.
This is expected as a lower viscosity leads to solutions that are harder to approximate with a linear space, and POD only produces
optimal approximation spaces in an~$\ell^2$-sense, not with respect to the maximum approximation error over all~$\nu$.
\par
We also report computational times and speedups in~\Cref{tab:pod_deim}:
While a single~FOM simulation requires around 19 minutes,
a simulation of the POD-DEIM~ROM requires less than 25 seconds,
resulting in a speedup factor of between 47 and 74, depending on the size of
the~ROM.
The reported total reductions times for computing the~ROM are dominated by the time required to compute
the snapshot data (``Number of snapshots''~$\times$ ``FOM simulation'').
Consequently, the MOR approach leads to substantial computational savings as soon as the total number of model
evaluations is larger than the number of snapshots required to build the~ROM.

\begin{table}[htbp]
    \centering
    \begin{tabular}{@{}c@{}}
       \includegraphics{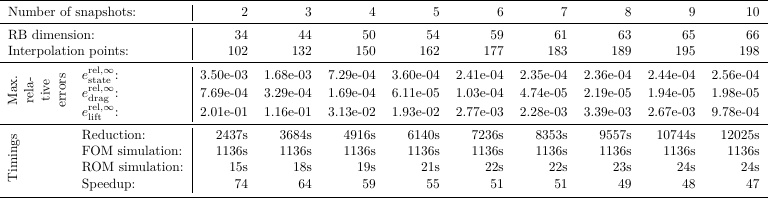}
    \end{tabular}
    \caption{Navier-Stokes example: POD-DEIM basis sizes, reduction errors~\Cref{eq:nsrelerrmax} and
    reduction times for different numbers of training snapshots.
    Reduction times include snapshot computation, basis generation and projection of the model.}
    \label{tab:pod_deim}
\end{table}

\subsubsection{POD-ML and POD-DEIM + POD-ML ROMs}\label{sec:navier_stokes_ml}
Next, we apply the POD-ML approach described in~\Cref{sec:ml_mor}.
We generate a POD basis for~6~FOM solution trajectories using the HAPOD algorithm as in \cref{sec:navier_stokes_pod_deim}.
As ML \PyClass{Regressor}, we compare kernel surrogates constructed via~VKOGA, deep neural networks using~\pyMOR's builtin~\PyTorch bindings and~Gaussian process regression from~\ScikitLearn.
The regressor is passed to~\PY{DataDrivenReductor} along with the parameters and the projection coefficients.
Here, we employ~\PY{DataDrivenReductor} instead of the~\PY{DataDrivenPODReductor} (see also~\Cref{sec:ml_mor}) to use a pre-computed
HAPOD basis~\PY{V} repeatedly for different~ML approaches.
We select~\PY{'solution'} as target quantity to inform the reductor that the provided data are state space coefficients rather than output data.
Since we consider a time-dependent problem, we further set the final time~\PY{T} and choose whether the~``time-vectorized'' strategy should be used or not.
In order to reconstruct a high-dimensional solution trajectory from the reduced coefficients, a linear combination with the reduced basis \PY{V} is formed.
The projection coefficients originate either from an orthogonal projection of FOM snapshots,
\begin{pythoncode}
    projection_coeffs = []
    for mu in training_mus:
        projection_coeffs.append(fom.solve(mu).inner(V))
        projection_coeffs = np.vstack(projection_coeffs)
\end{pythoncode}
or from solving the~POD-DEIM-ROM (\cref{sec:navier_stokes_pod_deim}):
\begin{pythoncode}
    projection_coeffs = []
    for mu in training_mus:
        projection_coeffs.append(rom_pod_deim.solve(mu).to_numpy().T)
        projection_coeffs = np.vstack(projection_coeffs)
\end{pythoncode}

For the~VKOGA and~GPR regressors, we use the~``time-vectorized'' ML strategy whereas for~DNNs, we use the~``random-access-in-time'' variant.
In order to understand these choices, consider the amount of training data available in the two settings and the input and output dimensions:
In the~``time-vectorized'' case, the input to the machine learning surrogate is solely the
parameter~$\mu\in\Params \subset \R$, whereas the output is the entire trajectory of reduced coefficients, resulting in an output dimension of~$n_t\cdot N$, where~$n_t=16,000$ denotes the number of time steps and~$N$ the dimension of the reduced space.
Learning such a high-dimensional output from a limited amount of $n_\mathrm{train}$ training parameters is not feasible for neural networks.
However, since all output components are treated separately by the kernel approximant with a diagonal vector-valued kernel, VKOGA and~GPR are indeed able to learn the required mapping.
In the~``random-access-in-time'' case, the input is again low-dimensional (parameter and time).
The output dimension, however, is only the reduced basis size~$N$, and
we have in total~$n_t\cdot n_\mathrm{train}\gg n_\mathrm{train}$ training data points, making the~``random-access-in-time'' approach suitable for training a neural network.
Since~$n_t\gg n_\mathrm{train}$, we subsample the time trajectories in this case to balance the parameter and time dimension in the input data.
\par
We consider the following scenarios in our experiments:
\begin{itemize}
    \item \FOMML: The output trajectories are directly learned from FOM output training data (setting \PY{target_quantity='output'}).
    \item \PODML: As described above, a POD basis \PY{V} is computed using the~HAPOD algorithm, FOM training trajectories are projected onto \PY{V} and the coefficients of the projections are approximated using ML regressors. The output trajectories are determined by applying the projected linear output functionals $gV$ to the reduced solution coefficients of the POD-ML ROM.
	\item \PODDEIMML: The~POD-DEIM ROM from~\Cref{sec:navier_stokes_pod_deim} is used to generate the training data. We can use more training parameters since the POD-DEIM ROM evaluations are significantly cheaper than~FOM solutions. The output trajectories are computed as in the \PODML case.
\end{itemize}

We perform a suitable scaling of the inputs and outputs of the machine learning surrogate.
For the input, we apply a logarithmic scaling followed by a min-max-scaling to the parameter and a min-max-scaling to the time component (if required).
The logarithmic scaling of the parameter is motivated by the logarithmic sampling of the parameter space,
emphasizing smaller viscosities, for which the approximation is more challenging.
The output values are scaled differently depending on the reduction approach:
For~VKOGA and~GPR, a simple min-max-scaling of outputs (\FOMML) or coefficients (\PODML and~\PODDEIMML) is sufficient, which uses the corresponding~\PY{MinMaxScaler} from~\ScikitLearn.
Interpolation surrogates such as~VKOGA appear to be less sensitive to scaling of the outputs since they interpolate (up to regularization) at the selected centers in any case.
Only the center selection might be affected by the scaling. For the neural networks, the outputs in the~\FOMML case are transformed as~$\tilde{y}=\ln(1+y)$, where~$y$ denotes a single component of the output followed by a min-max-scaling.
This choice is motivated by the strongly varying values of lift and drag when changing the parameter.
The coefficients for~\PODML and~\PODDEIMML are transformed as~$\tilde{y}=\operatorname{sign}(y)\cdot\ln(1 + \lvert y \rvert)$, where~$y$ refers to a single coefficient.
\par
Next, we provide further information on the parameters used for~VKOGA, neural networks and Gaussian process regression in the experiments:
\par
For~VKOGA, we apply~\pyMOR's builtin~Gaussian kernel with length scale~$0.3$ (the length scale has to be interpreted in the scaled parameter domain).
Moreover, we use the~\PY{'fp'}-greedy criterion for center selection, set the greedy tolerance to~$10^{-6}$ and choose a regularization parameter of~$10^{-12}$ to avoid overfitting of noisy data.
We do not restrict the number of centers that~VKOGA can select, but allow the algorithm to make use of all available training parameters as centers if necessary to reach the tolerance.
\par
For the neural networks, we employ a fully-connected neural network with three layers, consisting of~$128$ neurons each.
The Adam optimizer~\cite{kingma2015adam} is used for training with an initial learning rate of~$10^{-3}$ and a maximum of~$1000$ epochs.
We employ mini-batching with a batch size of~$2048$ and perform early stopping of the training if the loss does not decrease for~$250$ consecutive epochs.
The learning rate is adjusted based on a cosine annealing schedule using~\PY{CosineAnnealingLR} from~\PY{torch.optim.lr_scheduler}.
We found that restarting the training with different random initial conditions is not necessary with the aforementioned settings for batch size and number of epochs, which resulted in sufficient randomness during training.
As discussed above, we apply the~\PY{NeuralNetworkRegressor} in the~``random-access-in-time''-setting and perform a time subsampling by only selecting every~$10$th time step for generating training data.
\par
For GPR, we apply the same hyper-parameters as for~VKOGA, i.e.~a Gaussian kernel with shape parameter~$0.3$ and a regularization parameter of~$10^{-12}$.
In order to allow for a fair comparison with~VKOGA, we disable the hyper-parameter tuning of the Gaussian process regressor.
\par
For all three scenarios, \FOMML, \PODML and~\PODDEIMML, we conduct experiments using~VKOGA, DNNs and~GPR.
As training parameters in the~\FOMML and~\PODML cases, we use the same~$6$ parameters as for the~POD-DEIM ROM (\Cref{sec:navier_stokes_pod_deim}) and add~$24$ parameters sampled log-randomly from the parameter domain.
In the~\PODDEIMML case, we solve the~POD-DEIM ROM for~$200$ log-randomly sampled parameters.
The results are summarized in~\cref{tab:navier-stokes-ml-results}.

\begin{table}[htbp]
    \centering
    \begin{tabular}{@{}c@{}}
       \includegraphics{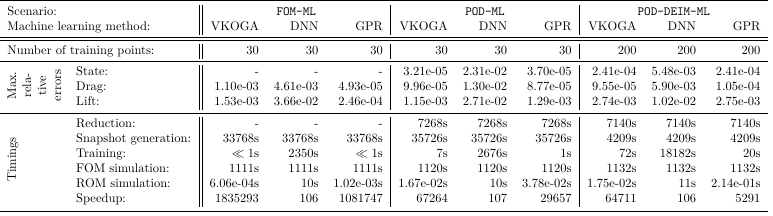}
    \end{tabular}
    \caption{Navier-Stokes example: Results of the machine learning-based methods.}
    \label{tab:navier-stokes-ml-results}
\end{table}

First of all, we observe that all methods presented in~\Cref{tab:navier-stokes-ml-results} reach relative errors in state, drag and lift of less than~$5\%$.
For the~\FOMML approach, VKOGA and~DNN reach similar errors for the drag, whereas the lift error of~VKOGA is an order of magnitude smaller compared to the~DNN.
The error of~GPR is two orders of magnitude smaller compared to~VKOGA for the drag and an order of magnitude smaller for the lift.
In particular, the speedup obtained by using~VKOGA is remarkable here with around two million compared to the~FOM simulation.
In the~\PODML approach also state predictions are available. Here we observe that the~DNN performs several orders of magnitude worse than~VKOGA and~GPR (which yield comparable errors).
The main reason is the limited amount of training parameters.
The~DNN has to predict a relatively large number of reduced coefficients and only~$30$ different parameter values are available.
In contrast, VKOGA and~GPR interpolate (up to regularization) the training data and then extend smoothly outside of the training parameters.
Considering the~\PODDEIMML approach allows us to collect much more training data by solving the~POD-DEIM~model instead of the~FOM.
In this example we use~$200$ training parameters at snapshot generation costs still smaller than for the~\FOMML and~\PODML approaches.
The~DNN benefits significantly from the larger amount of training data as shown by the errors in state, drag and lift all being almost an order of magnitude smaller compared to the~\PODML scenario.
For~VKOGA and~GPR in the~\PODDEIMML setting, we observe similar performance as for~\PODML, except for the state error, which is an order of magnitude larger using the reduced solutions instead of projections of~FOM solutions.
The speedup of~VKOGA is around~$60,000$ for both the~\PODML and the~\PODDEIMML case.
\par
In the following figures, we present the results for~VKOGA and~DNNs as machine learning surrogate.
We focus on these two methods to keep the figures lucid and since their implementation is provided by~\pyMOR.

\begin{figure}[htbp]
	\centering
	\includegraphics{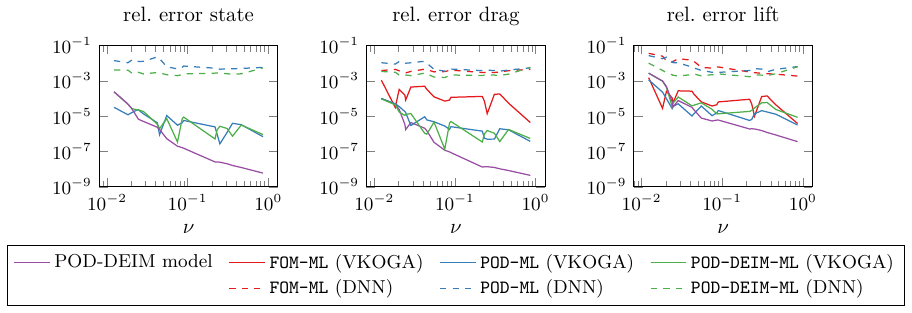}
	\caption{Navier-Stokes example: Error distribution in parameter space for the final models in all three scenarios using the maximum number of training parameters.}
	\label{fig:ml-errors-parameter-space}
\end{figure}
In~\Cref{fig:ml-errors-parameter-space}, we compare the error distribution of~VKOGA and~DNNs in the~\FOMML, \PODML and~\PODDEIMML scenarios over the parameter space for the models considered in~\Cref{tab:navier-stokes-ml-results}, i.e., using all training data available.
Further, the errors reached by the~POD-DEIM model are shown.
We observe that the errors for~VKOGA are more equally spread across the parameter domain compared to~POD-DEIM.
The trend of smaller errors for larger parameter values is only moderately visible.
Using~DNNs as surrogates leads to similar errors throughout the entire parameter set.
We further see that errors fluctuate significantly more for~VKOGA as machine learning method.
The relative errors obtained by~VKOGA in the~\PODML and~\PODDEIMML settings are on a similar level whereas the~\FOMML errors are slightly larger, as expected from the numbers stated in~\Cref{tab:navier-stokes-ml-results}.
\par
We also analyze the performance of the respective machine learning surrogates for different amounts of training data.
\begin{figure}[htbp]
	\centering
	\includegraphics{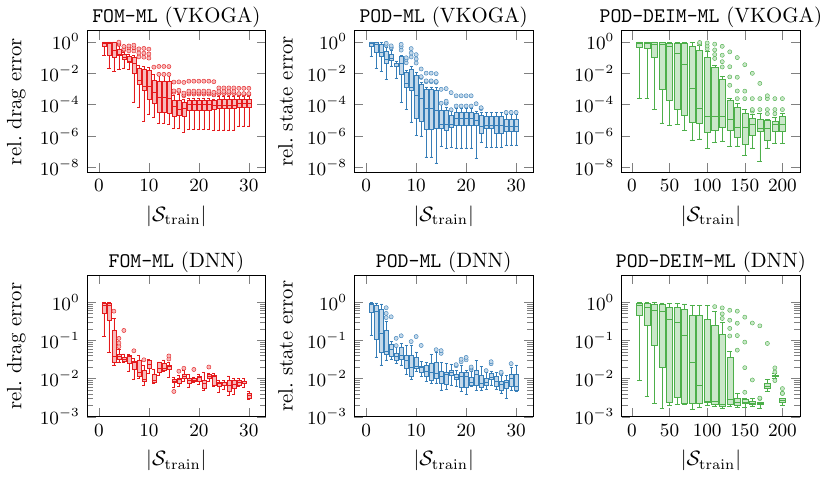}
	\caption{Navier-Stokes example: Statistics of relative error decay (drag for~\FOMML, state for~\PODML and~\PODDEIMML) for growing number of training parameters used to train~VKOGA- (top row) or~DNN- (bottom row) based~ML surrogates.}
	\label{fig:ml-errors-training-parameter-count}
\end{figure}
In~\Cref{fig:ml-errors-training-parameter-count} we show how the relative errors in outputs and state on the test set decrease with the number of training parameters.
For the~\FOMML case, we only show the drag errors here, the lift errors behave similarly.
For the~\PODML and~\PODDEIMML scenarios, mainly the state error is of interest since drag and lift are quantities derived from the computed reduced state.
For~VKOGA in the~\FOMML and~\PODML cases, we observe that the errors stagnate after~$15$ training parameters without substantial improvement.
Hence, already~$15$ parameters would have been sufficient for training.
Using~DNNs as surrogates for~\FOMML or~\PODML shows a drop in the error after having collected around~$3$ or~$4$ training parameters.
Adding more training data leads to slowly improving errors with no saturation reached for the final~$30$ parameters.
In the case of the~\PODDEIMML approach we see in the right plot of~\Cref{fig:ml-errors-training-parameter-count} that much more training data is required to reach a saturation point.
After around~$150$ parameters the test errors remain constant for~VKOGA and~DNNs.
We further highlight that the mean error when using~VKOGA in the~\PODDEIMML case is roughly two orders of magnitude smaller than the maximum error over the test set.
For~DNNs, all~$200$ training parameters are required in this example to also reduce the maximum error to be comparable to the mean error.

\begin{figure}[htbp]
	\centering
	\includegraphics{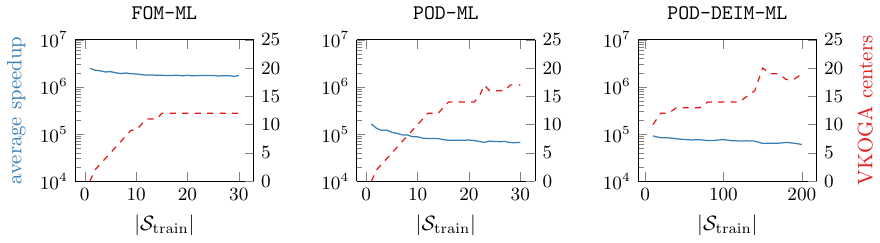}
	\caption{Navier-Stokes example: Speedup and number of selected centers with respect to the number of training parameters using~VKOGA as~ML surrogate.}
	\label{fig:ml-speedup-centers}
\end{figure}
Finally, we plot the speedup and the number of selected centers in~VKOGA in~\Cref{fig:ml-speedup-centers}.
In our experiments, not more than~$20$ centers are selected by the greedy algorithm in the~VKOGA training.
Therefore, in particular for the~\PODDEIMML scenario, it is useful to greedily select centers instead of performing kernel interpolation using all~$200$ training parameters.
A kernel expansion size of~$200$ would significantly reduce the speedup gained by the~VKOGA surrogate.
The plots further confirm that the number of selected centers has a slight influence on the average speedup.
However, the speedup is still remarkably large for the kernel-based surrogates taking also into account the accuracy of their predictions on the test set.
We remark that for the~\FOMML case, the number of selected centers plateaus when~$15$ or more training parameters are available, which matches the observation of stagnating errors from~\Cref{fig:ml-errors-training-parameter-count}.
For the~\PODML and~\PODDEIMML scenarios, the number of selected centers is not monotonically increasing with the number of training parameters.
For every training data count, the greedy center selection is run on the available training data from scratch, such that also some of the new training parameters might be selected early in the greedy procedure, leading to faster convergence of the algorithm.
The~VKOGA approaches further required no extensive hyper-parameter tuning or special scaling of outputs and coefficients in our experiments.
Consequently, for problems with smooth parameter-dependence, interpolation approaches such as~VKOGA might be preferable to neural networks, in particular when training data is limited.

\subsubsection{DMD}\label{sec:navier_stokes_dmd}
We finally consider a smaller viscosity of $\nu = 10^{-3}$, for which vortex shedding visibly occurs in
the solution (top row in \cref{fig:navier_stokes_solutions}).
To observe a time-periodic pattern, we set the inlet velocity to
\begin{align*}
	x(t)|_{\Gamma_\mathrm{in}} = \begin{cases}(t,0)^\T&\text{for }t\in[0,1],\\(1,0)^\T&\text{for }t>1,\end{cases}
\end{align*}
and restrict the solution to the time interval~$[6,8]$.

\begin{figure}[b]
	\centering
	\includegraphics{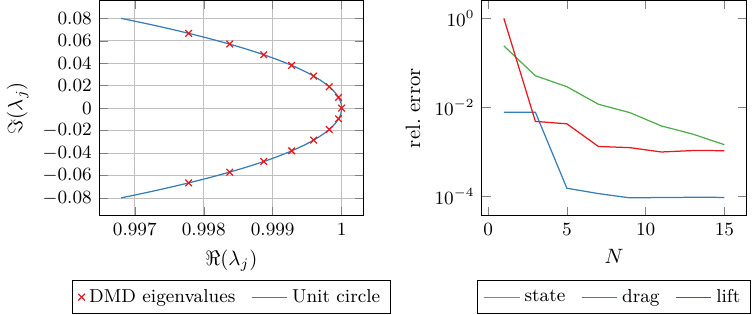}
	\caption{Navier-Stokes example: Discrete-time~DMD eigenvalues (left) and error decays in state reconstruction, lift and drag with respect to the number of~DMD modes (right).}
	\label{fig:dmd-omegas}
\end{figure}

We use values up to~$N=15$ for the number of retained~DMD modes in the reduced model and investigate the approximation error for different truncation ranks.
The resulting discrete-time~DMD eigenvalues and the relative errors for different numbers of~DMD modes are shown in~\Cref{fig:dmd-omegas}.
We observe that all eigenvalues lie (up to numerical errors) on the unit circle (highlighted in blue) in the complex plane.
This shows that we indeed consider the periodic regime.
Moreover, the relative errors in state, drag and lift decay with the number~$N$ of~DMD modes used for approximation.

\begin{figure}[htbp]
    \centering
    \includegraphics[width=\textwidth]{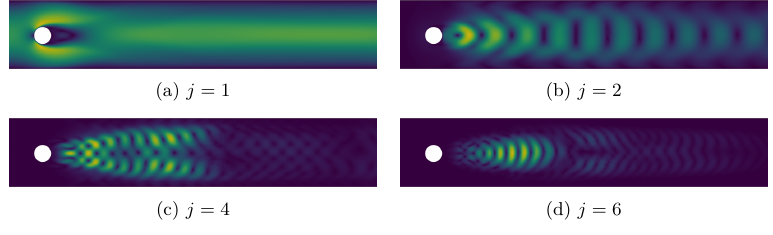}
    \caption{Navier-Stokes example: Magnitude of the first~4 velocity~DMD modes (there is one real mode followed by pairs of complex-conjugate modes; we depict one mode for each pair).}
    \label{fig:dmd-modes}
\end{figure}
\begin{figure}[tbp]
    \centering
    \includegraphics[width=\textwidth]{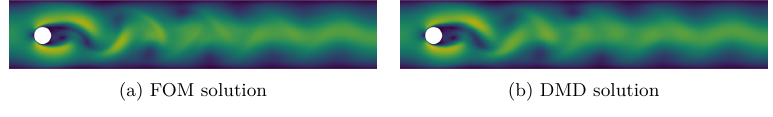}
    \caption{Navier-Stokes example: Magnitude of~FOM solution (left) and~DMD solution (right) velocity fields at the final time using~$N=5$~DMD modes.}
    \label{fig:fom-dmd-solution}
\end{figure}

The first DMD velocity modes are shown in~\Cref{fig:dmd-modes}.
The velocity~DMD modes exhibit typical spatial patterns, where the first mode is the mean flow and subsequent modes encode deviations from the mean.
The~FOM solution and the~DMD approximation for~$N=5$ are compared in~\Cref{fig:fom-dmd-solution}.
Using only~$N=5$ modes results in an accurate approximation with slight visual differences.
The main behavior of the flow until the end of the considered time interval is nevertheless captured accurately already with this low number of~DMD modes.
In the vicinity of the obstacle, the~DMD solution is slightly smoother compared to the~FOM solution.

\subsection{Mass-spring-damper chain}

As a second numerical example, on which we showcase system-theoretic~MOR methods,
we consider the mass-spring-damper chain example from~\cite{GugPGBS12}
of order~$200$ with two inputs and two outputs.
We choose it as it has equivalent forms both as a second-order system
\begin{equation}
    \label{eq:msd-so-fom}
    \begin{aligned}
        M \ddot{x}(t) + D \dot{x}(t) + K x(t) & = B u(t), \\
        y(t) & = B^\T \dot{x}(t),
    \end{aligned}
\end{equation}
and as a port-Hamiltonian system
\begin{equation}
    \label{eq:msd-ph-fom}
    \begin{aligned}
    E \dot{x}(t) & = (J - R) x(t) + G u(t), \\
    y(t) & = G^\T \dot{x}(t),
    \end{aligned}
\end{equation}
where
\begin{gather*}
    M = m I,
    \quad
    D = d I,
    \quad
    K =
    k
    \begin{bmatrix}
        1 & -1 \\
        -1 & 2 & -1 \\
        & -1 & 2 & -1 \\
        & & \ddots & \ddots & \ddots \\
        & & & -1 & 2 & -1 \\
        & & & & -1 & 2
    \end{bmatrix},
    \quad
    B =
    \begin{bmatrix}
        1 & 0 \\
        0 & 1 \\
        0 & 0 \\
        \vdots & \vdots \\
        0 & 0
    \end{bmatrix},
    \\
    E =
    \begin{bmatrix}
        K & 0 \\
        0 & M
    \end{bmatrix},
    \quad
    J =
    \begin{bmatrix}
        0 & K \\
        -K & 0
    \end{bmatrix},
    \quad
    R =
    \begin{bmatrix}
        0 & 0 \\
        0 & D
    \end{bmatrix},
    \quad
    G =
    \begin{bmatrix}
        0 \\
        B
    \end{bmatrix}.
\end{gather*}
We fix~$m = 4$, $d = 1$, $k = 4$ for the nonparametric case and
let~$d$ vary in the parametric case.

\subsubsection{Non-parametric case}\label{sec:msd-nonparametric}

First, we demonstrate model-based methods for non-parametric LTI systems
applied to the mass-spring-damper chain example,
specifically two unstructured (BT and IRKA) and
two structured (SOBTp and PH-IRKA) methods.
SOBTp is applied to the second-order
realization~\eqref{eq:msd-so-fom},
while PH-IRKA to the port-Hamiltonian realization~\eqref{eq:msd-ph-fom}.

\begin{figure}
    \centering
    \includegraphics{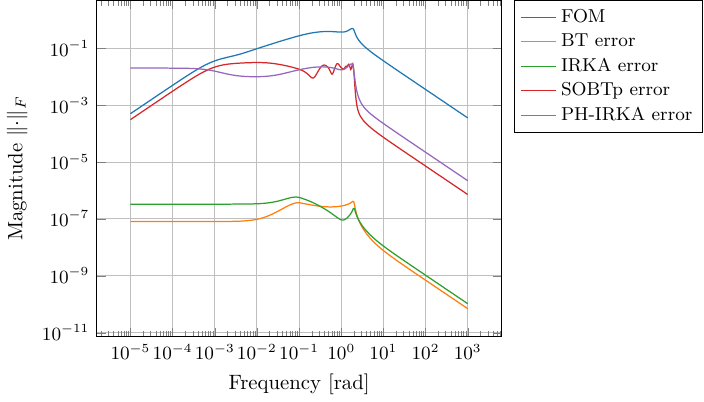}
    \caption{Mass-spring-damper chain example: Magnitude plots of the mass-spring-damper chain~FOM and
    error systems corresponding to four model-based ROMs of order $20$.}%
    \label{fig:msd-mag}
\end{figure}

\Cref{fig:msd-mag} shows the magnitude plots of the FOM and the four error
systems corresponding to ROMs of order $20$,
where for the SOBTp and second-order models we mean that a first-order
realization is of order $20$
(and a second-order realization is a system of $10$ differential equations)
to make the comparison fairer.
We observe that the FOM magnitude tends to zero for small and large frequencies,
and has a peak around 2~rad.
Next, we notice a significant gap between the unstructured and structured ROMs,
about 5 orders of magnitude,
while pairs of ROMs are close between themselves.
The gap was investigated in~\cite{BreU22}, where it was
found that applying PH-IRKA to a realization with a different Hamiltonian yield significantly
better results close to the unstructured IRKA.

\begin{figure}
    \centering
    \includegraphics{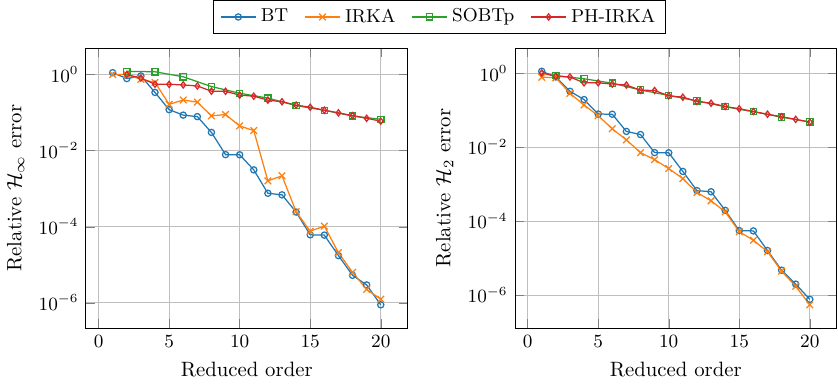}
    \caption{Mass-spring-damper chain example: Relative $\Hinf$ and $\Htwo$ errors for different reduced orders
    for four model-based methods.}%
    \label{fig:msd-order-error}
\end{figure}

To illustrate the results for other reduced orders,
\Cref{fig:msd-order-error} shows relative $\Hinf$ and $\Htwo$ errors for the
same four MOR methods across different reduced orders.
Again, we observe the gap between the unstructured and structured ROMs.
Furthermore, we notice bigger differences between BT and IRKA for lower orders
depending on the system norm.
As expected, IRKA performs better in terms of the $\Htwo$ norm.

\begin{figure}
    \centering
    \includegraphics{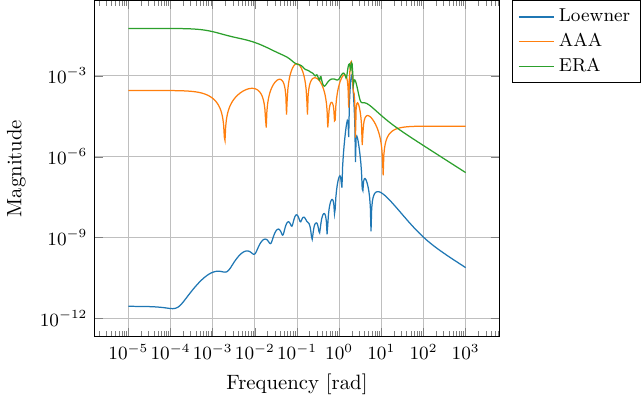}
    \caption{Mass-spring-damper chain example: Error system magnitudes for three data-driven ROMs of order $20$.}%
    \label{fig:msd-dd-tf-err}
\end{figure}

Next, we compare three data-driven methods: Loewner framework, AAA, and ERA.
For the Loewner framework and AAA, i.e., frequency domain methods,
we choose as data~$(i \omega_k, H(i \omega_k))$
for frequencies~$\omega = \mathtt{logspace}(-5, 3, 50)$.
In the case of ERA, we compute the impulse response of the FOM using implicit
Euler time stepping over the time interval $[0, 100]$ with step size of about
$9.8 \times 10^{-4}$, corresponding to the pole closest to the imaginary axis.
Then we subsample to obtain impulse response data with sampling time
$\Delta t = 0.05$.
\par
\Cref{fig:msd-dd-tf-err} compares the three methods for reduced order~$20$.
We see that they all have similar error around the FOM's peak,
with Loewner having the smallest error away from the peak.
AAA has a more uniform error, which is expected for a greedy method.
Note that AAA produces a proper, but not a strictly proper rational function
(there is a modification of AAA that enforces the result to be strictly proper,
which is not yet implemented in \pyMOR).
ERA has the worst performance for small frequencies,
which might be improved with a longer time horizon.

\begin{figure}
    \centering
    \includegraphics{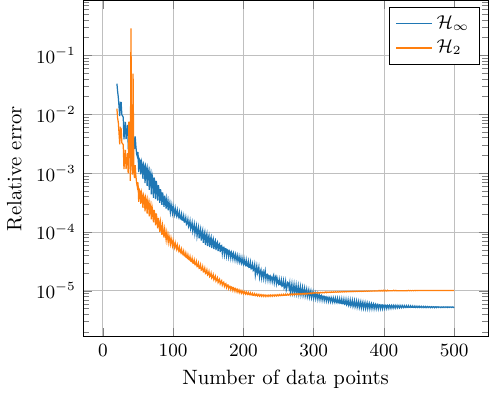}
    \caption{Mass-spring-damper chain example: Relative $\Hinf$ and $\Htwo$ errors of Loewner ROMs of order $20$
    for different numbers of data points $d$ and data from frequencies
    $\mathtt{logspace}(-5, 3, d)$.}%
    \label{fig:msd-loewner-data}
\end{figure}

We see that the data-driven methods perform worse than model-based methods,
which is expected since they only had access to the data and
not the whole model.
To see how the amount of data influences the results,
\Cref{fig:msd-loewner-data} shows relative errors for the Loewner ROMs of fixed
order $20$ but different amount of data.
We observe a convergence of the errors,
but still a difference of about 1 order of magnitude compared to BT and IRKA
as given in \Cref{fig:msd-order-error}.

\subsubsection{Parametric case}\label{sec:msd-parametric}

\begin{figure}
    \centering
    \includegraphics{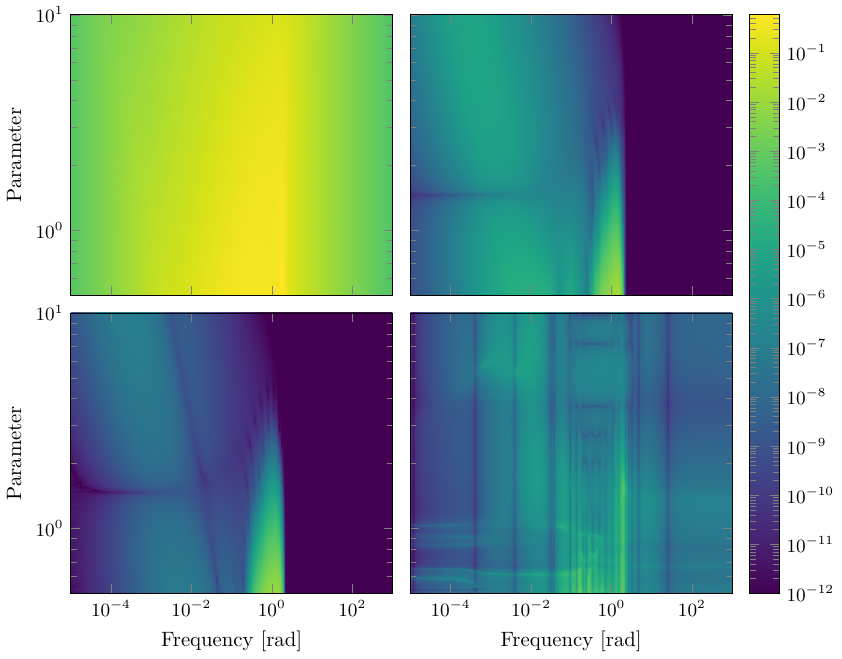}
    \caption{Mass-spring-damper chain example: Magnitude plots with respect to frequency and parameter of the
    parametric mass-spring-damper~FOM (upper left),
    parametric BT error (upper right),
    parametric IRKA error (lower left), and
    p-AAA error (lower right).}%
    \label{fig:msd-param-mag}
\end{figure}

We demonstrate~p-AAA on the parametric mass-spring-damper chain example
with the damping parameter $d$ varying in the interval~$[0.5, 10]$
and compare with parametric variants of BT and IRKA.
The magnitude of the~FOM transfer function can be seen in the upper left plot
of \Cref{fig:msd-param-mag}.
\par
For parametric BT and similarly for parametric IRKA,
we run the nonparametric method separately for each parameter value in
$\mathtt{geomspace}(0.5, 10, 10)$ for a fixed reduced order of $30$.
That way, we obtain $10$ local basis matrices $V_i$ with $30$ columns,
which we concatenate into a global basis matrix
$V_g =
\begin{bmatrix}
    V_1 & \cdots & V_{10}
\end{bmatrix}
$
with $300$ columns.
Then we use POD to truncate $V_g$ to $V$ and
apply Galerkin projection with $V$ to the port-Hamiltonian realization
to ensure asymptotic stability of the parametric ROM.\
The procedure using \pyMOR is illustrated in the following code snippet:
\begin{pythoncode}
    Vs = []
    for d in np.geomspace(0.5, 10, 10):
        reductor = ...Reductor(fom, mu=d)  # BT or IRKA reductor
        rom = reductor.reduce(30)
        Vs.append(reductor.V)
    V_g = cat_arrays(Vs)
    V_pod, _ = pod(V_g)
    V = V_pod[:N]
    pg = LTIPGReductor(fom, V, V)  # Petrov-Galerkin reductor
    rom = pg.reduce()
\end{pythoncode}
We found that truncating to $80$ and $70$ columns respectively for parametric BT
and IRKA gave similarly good results,
shown in \Cref{fig:msd-param-mag}.
The largest error is of order $10^{-2}$ around the frequency of 1~rad.
\par
To run p-AAA, as data we choose~$\mathtt{logspace}(-5, 3, 100)
\times \mathtt{geomspace}(0.5, 10, 100)$,
with corresponding frequency responses
(in total~$100 \times 100 = 10^4$ points).
The following code snippet shows how to create the respective reductor and
how to obtain the~ROM:
\begin{pythoncode}
    paaa = PAAAReductor([np.logspace(-5, 3, 100) * 1j, np.geomspace(0.5, 10, 100)],
                        phlti.transfer_function)
    rom_paaa = paaa.reduce()
\end{pythoncode}
We obtain a two-variate rational function of order $34$ in Laplace variable $s$
and order $17$ in the damping parameter $d$.
\Cref{fig:msd-param-mag} shows the result in the lower right plot.
We notice that the largest error is of order~$10^{-2}$ (similar to BT and IRKA)
and lines corresponding to interpolation points appear in the plot.

\section{Outlook and future plans}\label{sec:outlook}
In this paper, we have presented various algorithms and methods for MOR of parametric~PDEs and control systems implemented within~\pyMOR.
Particular focus has been on recent developments in the area of data-driven~MOR and how such approaches integrate and interact with classical model-based~MOR techniques.
We have highlighted that~\pyMOR provides a unified framework that enables model-based and data-driven MOR workflows for both classes of models.
\par
By means of numerical test cases, we showed~\pyMOR's capability of handling high-dimensional models implemented in external solvers, such as~\FEniCSx.
After setting up the~FOM as a \pyMOR \Model, the~MOR methods available in~\pyMOR can be used without significant additional effort.
Thus, \pyMOR makes it easy to compare a multitude of different MOR approaches for a given FOM and combine them into complex MOR pipelines.
\par
Current developments include the implementation of an adaptive hierarchy of model- or data-based ROMs as introduced in~\cite{haasdonk2023certified}.
For LTI control systems, there is ongoing work on quadrature-based (data-driven) balanced truncation~\cite{gosea2022datadriven} and an adaptive randomized version of ERA~\cite{minster2021efficient}.
\par
Furthermore, we are extending the available structure-preserving MOR methods, with a particular focus on port-Hamiltonian systems.
This includes the implementation of a structure-preserving version of~DEIM~\cite{chaturantabut2016structure}
and a port-Hamiltonian DMD variant~\cite{morandin2023porthamiltonian}.
Finally, support for bilinear control systems based on~\cite{benner2012interpolation} and~\cite{BenGR17} is planned.

\section*{Funding}
\begin{itemize}
    \item S.~Rave acknowledges funding by the Deutsche Forschungsgemeinschaft
        (DFG, German Research Foundation) under Germany's Excellence Strategy EXC 2044/2 --390685587,
        Mathematics Münster: Dynamics--Geometry--Structure.
    \item P.~Mlinari\'c acknowledges that the paper was supported by the
        European Union - NextGenerationEU through the National Recovery and
        Resilience Plan 2021--2026 Institutional grant of University of Zagreb
        Faculty of Science (Impact4Math).
\end{itemize}

\section*{Code availability} The source code used to carry out the numerical experiments presented in this contribution can be found at \url{https://github.com/pymor/2026_paper_code}~\cite{companion_code}.

\printbibliography

\end{document}